\documentstyle[12pt,epsf]{article}
\def\largelinestretch{\renewcommand{\baselinestretch}{1.3}}
\def\largelinestretch{\renewcommand{\baselinestretch}{1.3}}
\largelinestretch\small\normalsize

\title{
\bf BtoVVana: the package for analysis
  of  $B_s^0\to J/\psi\phi$ 
  and $B_d^0\to J/\psi K^*$  decays
}

\vspace*{-5mm}
\author{ S.~Shulga
\\[1ex]
\small
     Laboratory of Particle Physics, Joint Institute for Nuclear
     Research,
\hfill\\[-2mm]
\small
     141980 Dubna, Moscow region, Russia
\hfill\\[-2mm]
\small
     Francisk Skarina Gomel State University, Gomel, Belarus
\hfill\\[-3mm]
}
\date{}
\begin{document}

\thispagestyle{empty}
\begin{titlepage}
\thispagestyle{empty}
\maketitle
\begin{abstract}
  C++ package {\tt BtoVVana} is aimed to extract physics 
parameters in decays of the scalar neutral B-meson to two 
vector particles in the intermediate state, and two leptons 
and two pseudoscalar mesons in the final state.
  Improved angular moments method is implemented in the package {\tt BtoVVana}:
the method uses the time dependent
weighting functions and time-integrated observables
with variable upper time limit.
  These observables allow to keep full 
time informative contents of the decay.
  By using the package {\tt BtoVVana} it was shown 
that the statistical errors of extraction of
the observables strongly depends on the choice
of the angular weighting functions.
   The best angular weighting functions, which are
the linear combinations of the amplitude angular functions, 
are implemented in the package  {\tt BtoVVana}.
  The first version of the program contains algorithms 
to extract physics parameters in the case of untagged decays.
  The package {\tt BtoVVana} and program codes for the channels  
$B_s^0\to J/\psi\phi$ and  $B_d^0\to J/\psi K^*$
in the Monte Carlo generator {\tt SIMUB} were written
in parallel and mutually tested with high precision.
\end{abstract}
\newpage
\begin{center}
{\bf  Program summary }
\end{center}
{\bf\it Title of the program:} BtoVVana \\
{\bf\it Catalogue identifier:} XXXX\\
{\bf\it Program summary URL :} XXXX\\
{\bf\it Program obtainable from:} http://cmsdoc.cern.ch/$^{\sim}$\,shulga/BtoVVana/BtoVVana.html\\
{\bf\it Computer:} PC, two Intel 2.0 GHz processors, 512MB RAM \\
{\bf\it Operating system:} Linux Red Hat 6.1, 7.2, 7.3 and other
     platforms which allow to set ROOT package \\
{\bf\it Programming language used:} C++: gcc 2.96 or 2.95.2 compiler suite with g++\\
{\bf\it Size of the package:} 2.3 MB (0.4 MB compressed distribution archive), 
    without ROOT libraries (additional 120 MB)
    and without input files with events\\
{\bf\it Distribution format:} tar gzip file\\
{\bf\it Additional disk space required:} 
   Depends on the number of events: 
   35 Mb for 100\,000 events  (output of SIMUB generator)
{\bf\it Nature of the physical problem:}
    The package {\tt BtoVVana} has been developed
  to study the performance of the angular moments method of
  the $\Delta\Gamma$ determination from analysis
  of untagged decays 
  $B_s^0\to J/\psi(\to \mu^+\mu^-)\phi(\to K^+K^-)$.
    By using the package {\tt BtoVVana} it was shown that 
  the method of angular moments gives stable results and 
  is found to be an efficient and flexible tool 
  for measurements with $B_s^0\to J/\psi\phi$ 
  and $B_d^0\to J/\psi K^*$ decays.
\\
{\bf\it Method of solution:} 
  The method of angular moments allows to construct
a sequencial chain of extraction of the physics parameters
from events: $\Delta\Gamma$ $\to$ initial transversity 
amplitudes $\to$ weak CP-violating phase  $\to$ 
strong CP-conserving phases $\to \Delta M_s$.
  Standard angular moments method was improved by using 
the time dependent weighting function and time-integrated 
observables with variable upper time limit.
  These observables allow to keep full 
time informative contents of the decay.
  This improved method is used in the package {\tt BtoVVana}.
  It was shown that the statistical errors of extraction of
the observables strongly depends on the choice
of the angular weighting functions.
  The best  weighting functions were found and 
implemented in the package  {\tt BtoVVana}.
  The program {\tt BtoVVana} includes methods
of extractions of observables as independent modules.
  In frame of the package {\tt BtoVVana} 
user can include new analysis modules.
  The package {\tt BtoVVana} and program codes for the channels  
$B_s^0\to J/\psi\phi$ and  $B_d^0\to J/\psi K^*$
in the Monte Carlo generator {\tt SIMUB} were written
in parallel and mutually tested with high precision.
\\ 
{\bf\it Restrictions on the complexity of the problem:} 
 Program processes any number of files with any number of events; 
 program tested with about 200 files including 
 $2\times 10^6$ events
\\
{\bf\it Typical running time:}
  On a PC/Linux with 2.0 GHz processors BtoVVana package 
  spends 15 seconds for every 100 000 events. 
\vspace{10mm}
\hfill
\end{titlepage}

\def\largelinestretch{\renewcommand{\baselinestretch}{1.6}}
\largelinestretch\small\normalsize

\section{Introduction}
  The package {\tt BtoVVana} has been developed to analyze 
the decays of neutral pseudoscalar B-mesons into two vector 
mesons decaying into two muons and two pseudoscalar mesons.
  Two channels of this type are included in the B-physics
generator SIMUB~\cite{SIMUB} used here for precise testing 
of the package {\tt BtoVVana}:
$B^0_s(t),\overline{B}^0_s(t)\to J/\psi(\to \mu^+\mu^-)\,\phi (\to K^+K^-)$ 
and
$B^0_d(t),\overline{B}^0_d(t)\to J/\psi(\to \mu^+\mu^-)\,K^*(\to K\pi)$.
  These channels contain rich physics information because of
nontrivial angular distributions and time dependence of 
the decays.

  At present we have the information from D0 experiment 
at the TEVATRON about observation of 337 events with decay of the 
first type $B^0_s\to J/\psi \phi$ and 1370 events of the second 
type $B^0_d\to J/\psi K^*$~\cite{337evts}.
  The D0 data sample corresponds to an integrated luminosity 0.22 fb$^{-1}$,
collected in 2002-2004. 
 The statistics will be increased up to 4000 $B^0_s\to J/\psi \phi$ 
events at the D0 in the nearest future. 
 Assuming the $b\bar{b}$ cross section to be 100 $\mu$b,
the number of expected $B^0_s\to J/\psi \phi$ events, 
using 2 fb$^{-1}$ of integrated lumunosity, will be equal to 41400 
in the BTEV experiment~\cite{CERN-CKM}.
 About 8500 $B^0_d\to J/\psi K^*$ events are reconstructed in a date sample
taken by the BELLE detector in the KEKB $e^+e^-$ collider~\cite{8548evts}. 
 
  The systematic study of these decays will be performed after 2007 year 
on the LHC detectors.
  About 200\,000 $B^0_s\to J/\psi \phi$ events at the CMS, 
100 000 events at the ATLAS and 
75 000 events at the LHCb are expected to be obtained 
during first year of low luminosity operation~\cite{LHC_SMPhys}.

  Possibilities of the current experiments for
these important decays can be improved by using optimal
methods of the analysis.
  For large statistics the best method is likelihood fit
while for small statistics this method becomes unstable
for big number of unknown physics parameters,
and the angular moments method is more preferable ~\cite{dighe2}.
  The latest method allows one to build a sequencial chain 
of extraction of the physics parameters~\cite{SIMUB}.
  The goal of the paper is to formulate an optimal scenario
of this analysis which can be realized for real data
by using the package {\tt BtoVVana} presented here.
  Theoretical ideas realized in {\tt BtoVVana}
are based on papers~\cite{dighe2} and~\cite{SIMUB}.

 \section{ Structure of the package}

  Files of the package {\tt BtoVVana} are kept under the 
{\tt BtoVVana} directory which contains subdirectories {\tt mak, src, doc, res}:
directory {\tt src} contains the source codes of the program;
{\tt mak} contains {\tt Makefile}, command files 
          for compilation {\tt make\_release} and execution {\tt run};
{\tt bin} contains the results of the compilation (object files) 
          and the executable file (this directory is created authomatically
	  by command {\tt make\_release});
{\tt doc} includes documentations;
{\tt res} is the user directory for data and results.

  The structure of the directory tree one can be found in {\tt Makefile}:
{\small
\begin{verbatim}
source_dirs =../mak
source_dirs+=T_run_Read
source_dirs+=T_run_Read/T_Accum_Measure
source_dirs+=T_run_Read/T_Accum_Measure/T_AngMomMethod
source_dirs+=T_run_Read/T_Accum_Measure/T_AngMomMethod/T_WeightFunc
source_dirs+=T_run_Read/T_Accum_Measure/T_AngMomMethod/T_StandardModel_DG
source_dirs+=T_run_Read/T_Accum_Measure/T_AngMomMethod/T_WeightFunc\
            /T_TimeIntObs
source_dirs+=T_run_Read/T_Accum_Measure/T_AngMomMethod/T_WeightFunc
            /T_TimeIntObs/T_Physics
source_dirs+=T_Utility
\end{verbatim}
}
  The directory tree reflects a logical structure and dependences
of the classes of the package.  
  Command {\tt make\_release} processes all {\tt *.C} files
in directories included in the list above.
  As a rule the name of the directory coincides with 
the name of the class which is placed in the directory.
  The structure of the program is shown in Fig.1.

  Class {\tt T\_run\_Read} is intended for reading 
the events from external files.
  This class contains the main loop over events.
  For mode {\tt Mode\_Loop = 2} the loop calls 
the virtual dummy method which will be overlapped
in the derived class. 
  Class {\tt T\_run\_Read} contains the methods 
for regime {\tt Mode\_Loop = 3} in which the angle 
and time distributions are created (see Fig. 3).

  The daughter class {\tt T\_Accum\_Measure} overlaps some 
dummy methods of the mother class {\tt T\_run\_Read} and 
is intended for regime {\tt Mode\_Loop = 2}.
  The main data members of the class are the so called "containers"
which acummulate the information event by event, average the observables,
and write the results to the final listing.

  The class {\tt T\_AngMomMethod} carries out the extraction 
of the observables by using the angular moments method.
  Two "containers" are impemented in the class {\tt T\_Accum\_Measure}:
both of the  "containers" are described as objects
of the same class {\tt T\_AngMomMethod} with different 
angular weighting functions.

  The main data members of the class  {\tt T\_AngMomMethod} are
four objects of the class {\tt T\_TimeIntObs}
and three objects of the class {\tt T\_StandardModel\_DG}.

 The class {\tt T\_TimeIntObs} contains descriptions of 
the time-integrated observables which are obtained 
by accumulation of the event weights.

 The class {\tt T\_StandardModel\_DG} performs calculations
of the width $\Gamma$ and width difference $\Delta\Gamma$ 
by using the found observables in case of 
the Standard Model prediction of the small CP-violating 
weak phase.


  The weights of the events are found by using the class 
{\tt T\_WeightFunc} which is the mother class for class
{\tt T\_TimeIntObs}.

  Extraction of the final observables based on some predefined 
physics parameters with its errors which described in class 
{\tt T\_Physics}.
  The object of this class should be created in the {\tt main} 
program by the user in case of processing the real data.
  In case of processing the data from generator {\tt SIMUB}
one needs to set logical parameter {\tt bRealData = false} 
in the {\tt main} function.
  In this case the program will read the generator physics 
parameters  from the first event.
  
  Directory {\tt T\_Utility} includes some auxiliary functions
and classes.  
    
  The structure of the package {\tt BtoVVana} 
is the same as the structure of the generator {\tt SIMUB}~\cite{SIMUB}
which allows to obtain the events 
$B^0_s(t),\overline{B}^0_s(t)\to J/\psi (\to l^+l^-)\,\phi (\to K^+K^-)$
and
$B^0_d(t),\overline{B}^0_d(t)\to J/\psi(\to l^+l^-)\,K^*(\to K\pi)$ 
with full physics contents and high precision
of generation of the kinematical variables.

 \section{ Preparation of input files by the {\tt\bf SIMUB} generator}
  Subpackage {\tt SIMUB/BB\_dec} allows to obtain 
the events with $B_s^0\to J/\psi\phi$ 
and $B^0_d\to J/\psi K^*$ decays
with full physics contents by setting the channel option 
"{\tt COPT SI1}"
in command file {\tt BB\_dec/mak/run}~\cite{SIMUB}
and using the decay mechanism option "{\tt B0SM 1}" 
("{\tt B0DM 1}") and decay channel option "{\tt B0SC  1}" 
("{\tt B0DC  1}") for $B^0_s$ ($B^0_d$) mesons.
  The format of the {\tt SIMUB} event is defined by option {\tt FRMT}.
  Subpackage {\tt SIMUB/BB\_dec} has a mode {\tt Mode\_Loop==2 } 
to produce events in the format readable by 
the program {\tt BtoVVana}.
  The variable {\tt Mode\_Loop} is set in the file 
{\tt BB\_dec/mak/run} by option {\tt MODD 2}.
  In this case the format defined by option {\tt FRMT} 
is ignored and the events are written in {\tt ROOT} tree 
in the format defined in the constructor of class {\tt T\_Loop}
(subpackage {\tt SIMUB/BB\_dec}) in the following way:
{\small
\begin{verbatim}
if(Mode_Loop==2){ 
 hfile= new TFile(fname,"RECREATE","B0s->J/PsiPhi or B0d->J/PsiK* tree");
 tree = new TTree("T","B0s->J/PsiPhi decay tree");
 tree->Branch("VB" ,&VB ,"x/D:y:z:t:tau");
 tree->Branch("PB" ,&PB ,"x/D:y:z:E:M");
 tree->Branch("Pa" ,&Pa ,"x/D:y:z:E:M");
 tree->Branch("Pa1",&Pa1,"x/D:y:z:E:M");
 tree->Branch("Pa2",&Pa2,"x/D:y:z:E:M");
 tree->Branch("Pb" ,&Pb ,"x/D:y:z:E:M");
 tree->Branch("Pb1",&Pb1,"x/D:y:z:E:M");
 tree->Branch("Pb2",&Pb2,"x/D:y:z:E:M");
 tree->Branch("DFG",&DFG,"delta1/D:delta2:Wphi:nA02:nAP2:nAT2:
    GB:GL:GH:DG:MB:ML:MH:DM:dx:dy:dz:dt:Tmax:cThb1:cTha1:sChi:
    cChi:Chi:Blifetime:fDist:KFCode/I");}
\end{verbatim}
}
  Here the {\tt fname} is a name of the output file defined by option 
{\tt OUTF 'NAME.root'},
the branches {\tt VB} and {\tt PB} define the secondary 4-vertex 
  of $B$-decay and  4-momentum of B-meson,
branches  {\tt Pa} and {\tt Pb}  define 4-momentum of 
  $J/\psi$ and $\phi$ ($K^*$) mesons, 
branches {\tt Pa1} and {\tt Pa2} define 4-momentum
  of $\mu^+$ and $\mu^-$ mesons,
branches {\tt Pb1} and {\tt Pb2}  define 4-momentum 
  of $K^+$ and $K^-$ mesons ($K$ and $\pi$ for $K^*$), 
the branch {\tt DFG}  defines the physics parameters 
  of the decay $B\to J/\psi\phi$ ($B^0_d\to J/\psi K^*$): 
  strong CP-conserving phases $\delta_1 = \mbox{arg}(A_{||}^*A_{\bot})$ 
  (named as {\tt delta1}) and 
  $\delta_2 = \mbox{arg}(A_{\bot})$  
  ({\tt delta2}; we set $\mbox{arg}(A_0) = 0$), 
  weak CP-violating phase $\phi$ ({\tt Wphi}), 
  initial transversity amplitudes squared 
  $A_0^2,\, A_{\bot}^2,\, A_{||}^2$ (names {\tt nA02, nAP2, nAT2}), 
  $B_s^0$ ($B^0_d$) width $\Gamma$,  
  widths of light and heavy states $\Gamma_L,\,\Gamma_H$,
  $\Delta\Gamma = \Gamma_H-\Gamma_L$ (with the corresponding 
  program names {\tt GB, GL, GH, DG}),
  $B$ mass  $M_{B^0}$ and masses of light and heavy states 
  $M_L, M_H$, $\Delta M \equiv M_H - M_L$ 
  (program names {\tt MB, ML, MH, DM}),
  generator resolutions~\cite{SIMUB} or detector resolution 
  named by  {\tt dx, dy, dz, dt}
  for four kinematical variables $\cos\Theta_{\mu^+}$, 
  $\cos\Theta_{K^+}$, $\chi$ and $t$, correspondently,
  where three angles are defined in Fig.~2,
  and $t$ is the $B$ meson proper lifetime.

  If the program {\tt BtoVVana} is used for the generator 
data, then the data members from the branch 
{\tt DFG} contain physics parameters from the generator.
  In case of analysis of real data these data members 
should contain theoretical predictions to calculate some 
values which depend on the measured values and external 
parameters and for comparison.

  Any number of the input files with events 
can be treated by the program {\tt BtoVVana}.
The names of these files are written in the main program.

 \section{ Quick start with the package {\tt BtoVVana}}
  The package is tested on Linux (RedHat 7.x, 8.x) platforms
and uses the {\tt ROOT} package~\cite{ROOT}.
  To run the package, it is necessary to set the 
environment variables of the {\tt ROOT}.

   To install the program, it is enough to untar the file
{\tt BtoVVana.tar.gz} in the working directory 
and the {\tt BtoVVana} directory with the structure
described above will be created.

  The main function is placed in the file 
{\tt mak/T\_Accum\_Measure\_main.C}.
  The user can set here the following parameters:
\begin{itemize}  
 \item  code of B-meson {\tt KF\_Code} 
 (Pythia KF-code is 531 for $B^0_s$ and 511 for  $B^0_d$);
 \item number of the input files {\tt NFiles} and its 
      names collected in the array {\tt cFile};
 \item {\tt Mode\_Loop} = 2 to extract the physics parameters
   or {\tt Mode\_Loop} = 3 to draw the angular and proper time
   distributions;
 \item maximal proper lifetime used to select the 
       events (or for Monte Carlo generation in generator) 
       {\tt Time\_maximal\_mmDc};
 \item the upper limit for the proper lifetime of B-meson 
      {\tt Time\_UP\_mmDc} (in units of [mm/c]), 
      {\tt Time\_UP\_mmDc} = {\tt Time\_maximal\_mmDc} as a rule;
 \item second choice of the upper limits for proper lifetime 
       {\tt Time\_Zero\_mmDc};
 \item parameter {\tt GammaS\_mmDcm1} may be set as arbitrary;
    to minimize the errors of measurements,
    one needs to set {\tt GammaS\_mmDcm1} closely to the true
    value of $\Gamma$ 
    ($\Gamma$ is B-meson width in units [mm/c]$^{-1}$);
 \item logical parameter {\tt bRealData = true} in case of 
   analysis of real data, otherwise the program will process
   the events from the generator; the only difference is that
   in case of {\tt bRealData = false} the user does not need 
   to define the object of {\tt T\_Physics *Phys} 
   in the {\tt main} finction because this object will be 
   defined within the program as the data member by method 
   {\tt Parameter\_SIMUB} of the class  {\tt T\_AngMomMethod} 
   by using the special physics records in the events written 
   by generator {\tt SIMUB}
   (see branch {\tt DFG} in the class {\tt T\_run\_Read} which 
    one can found in package {\tt BtoVVana} 
    and in generator {\tt SIMUB});
   in case of {\tt bRealData = true} the user needs to set 
   an object of the class {\tt T\_Physics} in the {\tt main} function 
   by using the data from the previous run of the program or from
   other sources;
 \item number of events {\tt NEVENTS} (will be ignored if
    it is less than the number of the events in all the input files);
 \item to draw a simple histogram, one has to set {\tt bSIMPLE\_HISTOGR = true}
   otherwise it will be drawn histograms with error bars
   as it is shown in Fig.3;
 \item logical variable {\tt DEBUG = true} for extended information
       in the listing used for debug of the program;
 \item logical variable {\tt bShortINF = true} for short 
       information in the listing;
\end{itemize}  

  To compile the program, one has to go into {\tt mak} directory 
and execute the command {\tt make\_release}.  
  After that, one should start the program by the command {\tt run}.

  Example of the main program and listing with results of the program
are placed in section "Test Run Input and Output".

 \section{Class {\tt\bf T\_run\_Read}: loop over the events}
  The main part of the class {\tt T\_run\_Read} was done
authomatically by the {\tt ROOT} method {\tt MakeClass} 
of the {\tt TTree} class.
  The constructor of the class {\tt T\_run\_Read} 
contains the following parameters:
{\tt KF\_Bmeson} is a PYTHIA KF-code of B-meson,
{\tt NumbFiles} is  a number of the files for analysis, 
{\tt Mode\_Loop} is a processing mode, and 
{\tt cFile[]} is a list of the file names.

  Class {\tt T\_run\_Read} contains the loop which reads 
the data and writes the event information in the folowing 
data members of the class:
{\small
\begin{verbatim}   
double VB_x,  VB_y,  VB_z,  VB_t, VB_tau, // Vertex and lifetime
  PB_x,  PB_y,  PB_z,  PB_E,  PB_M,     //    B momentum and mass
  Pa_x,  Pa_y,  Pa_z,  Pa_E,  Pa_M,     //J/psi momentum and mass
 Pa1_x, Pa1_y, Pa1_z, Pa1_E, Pa1_M,     //  mu+ momentum and mass
 Pa2_x, Pa2_y, Pa2_z, Pa2_E, Pa2_M,     //  mu- momentum and mass
  Pb_x,  Pb_y,  Pb_z,  Pb_E,  Pb_M,     //  Phi/Ktar momentum and mass
 Pb1_x, Pb1_y, Pb1_z, Pb1_E, Pb1_M,     //  K+(K)  momentum and mass
 Pb2_x, Pb2_y, Pb2_z, Pb2_E, Pb2_M;     //  K-(pi) momentum and mass
   // Physics, kinematics and Monte Carlo parameters (branch DFG):
double DFG_delta1,DFG_delta2,// strong CP conserving phases
 DFG_Wphi,                     // weak CP violating phase
 DFG_nA02,DFG_nAP2,DFG_nAT2,   // initial polarized amplitudes squared (0,||,T)
 DFG_GB  ,DFG_GL,DFG_GH,DFG_DG,// see definition in previous section 
 DFG_MB,DFG_ML,DFG_MH,DFG_DM,  // see definition in previous section
 DFG_dx,DFG_dy,DFG_dz,DFG_dt,  // see definition in previous section
 DFG_Tmax,                     // maximal alowed B lifetime 
 DFG_cThb1,DFG_cTha1,          // cos(Theta_K+) cos(Theta_mu+)
 DFG_sChi, DFG_cChi,DFG_Chi,   // cos(Chi),sin(Chi), Chi
 DFG_Blifetime,       // B Lifetime
 DFG_fDist;           // value of distribution function for given 
                      // cos(Theta_K+),cos(Theta_mu+), Chi, B-lifetime
 int DFG_KFCode;      // PYTHIA KF code of B meson
\end{verbatim}
}
  Using the method {\tt Mom\_To\_LorentzVector}
the event momenta and vertex are rewritten 
in {\tt ROOT} Lorentz vector variables:
{\tt P\_B\_L, P\_a\_L,P\_a1\_L, P\_a2\_L, P\_b\_L, P\_b1\_L, P\_b2\_L,
V\_B\_gen, V\_B\_dec, DVertex\_B = V\_B\_dec - V\_B\_gen}.
  The data members - {\tt cThe\_b1\_R}, 
{\tt cThe\_a1\_R}, {\tt Chi\_R}, {\tt B\_lifetime\_R } -
present the values $\cos\Theta_{\mu^+}$,
$\cos\Theta_{K^+}$, $\chi$, $t$ (see Fig.2) which are
calculated for the current event by the method
{\tt Reco\_Helicity\_Angles}
(if the position of a method is not given,
 then the method belongs to the class {\tt T\_run\_Read}
in this section by default).

  The number of the input files for the processing is set in 
the constructor of the class {\tt T\_run\_Read} by parameter {\tt N\_umbFiles}.
 The file names are collected in the character array 
in the main program and passed in the constructor
by the parameter {\tt cFile[]}.
 
  The method {\tt Loop(NEVENTS)} performs the loop over 
all the events in the file chain.
 The user can find two logical keys {\tt bSIMUB\_Angles} 
and {\tt bSIMPLE\_HISTOGR} inside the method {\tt Loop}.
 If {\tt bSIMUB\_Angles = true}, then 
the angles  and B lifetime directly from generator 
{\tt SIMUB} (from the branch {\tt DFG})
will be used for histograms, otherwise the angles and B lifetime 
which were reconstructed from the momenta and vertex, are used
for histograms;
  if {\tt bSIMPLE\_HISTOGR = true}, then we can obtain the pictures
with simple histograms for the three angles and B lifetime,
while for {\tt bSIMPLE\_HISTOGR = false} one can obtain
histograms with error bars.
  The latest kind of pictures for decay $B_s^0\to J/\psi\phi$
can be seen in Fig.~3 for 100\,000 events.
  We have used the  "main setting" of physics parameters
by default for  $B_s^0\to J/\psi\phi$($B_d^0\to J/\psi K^*$)
decays:
$\delta_1  = \pi$, $\delta_2 =0$, $\phi^{(s)}_c = 0.04$, 
$A_0^2      = 0.54(0.56)$, 
$A_{\bot}^2 = 0.16$, $A_{||}^2   = 1-A_0^2-A_{\bot}^2$, 
$\Gamma = 1/\tau_B$, 
$\tau_B = 1.464 (1.571)\,\,\mbox{ps}$, 
$\Delta\Gamma/\Gamma = -0.2 (-0.01)$,
$T_{max} = 2 \mbox{mm/c}$, $N_{reso} = 50\,000$, 
$dt=dx=dy=2/N_{reso}, dz = 2\pi/N_{reso}$,
$		T_{max} = 2\,\frac{mm}{c}$, 
$x \equiv \Delta M/\Gamma = 20 (0.73)$,
where the values for $B_d^0\to J/\psi K^*$ are shown in brackets
if they do not coincide.

  Switch of regimes {\tt Mode\_Loop} are placed in the main program.
  Class {\tt T\_run\_Read} includes methods in 
a simple mode {\tt Mode\_Loop = 3}.
  Methods for the {\tt Mode\_Loop = 2} are placed in the derived class
{\tt T\_Accum\_Measure} (see the next section).
  All other values of {\tt Mode\_Loop} are not used in the current 
version of the program.

  It is convenient to create a separate derived class {\tt T\_Accum\_Measure}
specially  to treat the events.

  The main loop over the events is placed in class 
{\tt T\_run\_Read} with non virtual definition and 
therefore it is not redefined
in the derived class {\tt T\_Accum\_Measure}.
  The program uses the same method {\tt Loop}
placed in the parent class {\tt T\_run\_Read} 
for all regimes.

  The method  {\tt Loop} contains the switch on different 
regimes of processing by the key {\tt Mode\_Loop}.
  This switch has the following form (it is a simplifyed form 
of the loop):
{\small
\begin{verbatim}   
  for(jentry=0; jentry<NEVENTS ; jentry++) {
    if(Mode_Loop==2)
     if(!Reco_Accum_Measure()) out("$WARNING 1(goto next evt)",0);
    else if(Mode_Loop==3) Fill_Hist_RECO(); }
  if     (Mode_Loop==2) Average_Measure_OutResult();
  else if(Mode_Loop==3) Draw_Hist(0);
\end{verbatim}
}
  Within the class {\tt T\_run\_Read} there are
dummy virtual methods {\tt Reco\_Accum\_Measure} and 
{\tt Average\_Measure\_OutResult}
which are overlapped in the derived class. 
  The method {\tt Reco\_Accum\_Measure} transforms
the momenta of the particles into angles 
and accumulats observables.
  The method {\tt Average\_Measure\_OutResult} performs 
averaging and printing the results of the observables extraction.

 \section{Class {\tt\bf T\_Accum\_Measure}: accumulation of the observables}

  Class {\tt T\_Accum\_Measure} has  {\tt T\_run\_Read} 
as a parent class and its  constructor has the form:
{\small
\begin{verbatim} 
T_Accum_Measure(bool DEBUG,
   int KF_B_To_Analyse, int NumbFiles,  char *cFile[],
   int Mode_Loop, double TimeUP_mmDc, double Time_maximal_mmDc, 
   double Time_Zero_mmDc, double GammaS_mmDcm1,
   bool bRealData, T_Physics *Phys);
\end{verbatim}
}
\hspace{-7mm} where the parameters {\tt DEBUG}, {\tt KFB\_To\_Analyse}, 
{\tt NumbFiles}, {\tt cFile[]} and {\tt ModeLoop}
are used for the constructor of the parent class  {\tt T\_run\_Read}
described in the previous section.
  All other parameters are described in  section 4.
  
  At present the class {\tt T\_Accum\_Measure}
includes the extraction of observables
by the angular moments method described in~\cite{SIMUB}.

  The main methods of the class overlap the parent virtual 
methods {\tt Reco\_Accum\_Measure} and {\tt Average\_Measure\_OutResult}.
  These methods are called by derived method
{\tt T\_run\_Read::Loop} which is not overlapped in the class
{\tt T\_Accum\_Measure}.
  Class {\tt T\_Accum\_Measure} has the "containers" to
accumulate the information in the loop over the events.
  These data are processed and printed in their final form
in function {\tt Average\_Measure\_OutResult}.
  "Containers" are represented as the following data members of the class:
{\small
\begin{verbatim} 
  T_AngMomMethod *AngMomMethod_A;
  T_AngMomMethod *AngMomMethod_B;
\end{verbatim} 
}

\vspace{-5mm}
\hspace{-7mm} where {\tt AngMomMethod\_A}({\tt AngMomMethod\_B}) 
are used to treat the events by using the angular moments method 
with the weighting function which is described in~\cite{SIMUB} 
as "set A"("set B").

  The user can introduce new "containers" for other ways of treatment.
  For example, tagged events of decay 
$B_s^0(t)\overline{B}_s^0(t)\to J/\psi\phi$
containing the oscillation phenomenon allow to extract the parameter 
$\Delta M$.
  According to the scenario  presented here,
the extraction of the $\Delta M$ by using the tagged samples
should be done after analysis of the untagged sample.
  The method to extract $\Delta M$ and corresponding "container" 
will be developed in the next version of the program  {\tt BtoVVana}.
  
  A pointer to the object of the class 
{\tt T\_Accum\_Measure *AM} is defined in the main function.
  Constructor of the class {\tt T\_Accum\_Measure} 
defines the "containers" (parameters of the constructor 
will be described in the following section):
{\small
\begin{verbatim} 
 AngMomMethod_A = new T_AngMomMethod(1, ...);
 AngMomMethod_B = new T_AngMomMethod(2, ...);
\end{verbatim} 
}
  Then the main program calls the loop method 
inherited from the parent class {\tt AM $\to$ Loop(500000)}.
  It passes the number of events to the loop to be processed.
  The events from the external files are read and written
into the data members of the parent class  {\tt T\_run\_Read}.
  The structure of the loop was shown in the previous section.
  As you can see, the loop calls the method {\tt  Reco\_Accum\_Measure}
which is overlapped in the class {\tt T\_Accum\_Measure} 
and has the following form:
{\small
\begin{verbatim} 
bool T_Accum_Measure::Reco_Accum_Measure() {
  Mom_To_LorentzVector(); 
  Reco_Helicity_Angles();
  Count_MeasureEv++; if(Count_MeasureEv==1) Measure_Init();
  Measure_Accum(); return true; }
\end{verbatim} 
}
  The first method {\tt Mom\_To\_LorentzVector} transforms
the momenta to the {\tt ROOT} Lorentz vectors and to
the angles (in the frame shown in Fig.~2) and the B-meson proper 
lifetime calculated by using the function 
{\tt Reco\_Helicity\_Angles}.
  The last function is placed in directory  {\tt T\_Utility/F\_Reco\_HelicityAngles}.

  Before treatment of the first event the initialization is performed
by means of the method {\tt Measure\_Init}.
  In the treatment of events from the generator {\tt SIMUB} 
the method {\tt Measure\_Init} calls the method 
{\tt Parameter\_SIMUB} to initialize the "containers" 
{\tt AngMomMethod\_A} and {\tt AngMomMethod\_B} by the 
physics parameters (in this section the methods without 
references have the placement in the class {\tt T\_Accum\_Measure} 
by default).

  These parameters are used to extract the observables from 
the time-integrated observables which have been obtained by
class {\tt T\_TimeIntObs}, and to perform  analytical 
calculations to compare them with the approximate 
estimations obtained by the Monte Carlo method 
(see description of the class  {\tt T\_Physics} below).
  It is helpful to test the both the event generator 
and the analysing package {\tt BtoVVana}.

  As it is described in the previous section, 
the method {\tt Loop} derived from class  {\tt T\_run\_Read}
calls the virtual method {\tt Reco\_Accum\_Measure} which 
overlapped in the class described in this section.
 Further the chain of calls is the following:
{\tt Reco\_Accum\_Measure()} $\to$ 
{\tt Measure\_Accum()} $\to$ {\tt Measure\_Accum\_AngMomMethod()}.
  The structure of the latest method is very general:
{\small
\begin{verbatim} 
void T_Accum_Measure::Measure_Accum_AngMomMethod(){
 AngMomMethod_A->Init_Angles(cThe_b1_R, cThe_a1_R, Chi_R, B_lifetime_R);
 AngMomMethod_A->Accumulation_TimeIntObs  ();
 AngMomMethod_A->Set_False_FlagInit_Angles();
 AngMomMethod_B->Init_Angles(cThe_b1_R, cThe_a1_R, Chi_R, B_lifetime_R);
 AngMomMethod_B->Accumulation_TimeIntObs  ();
 AngMomMethod_B->Set_False_FlagInit_Angles();
}
\end{verbatim} 
}
  It is not difficult to include other user "containers" 
into this method.

  In method {\tt T\_run\_Read::Loop(int)} after the loop
the method \\
{\tt Average\_Measure\_OutResult} is called.
 This method performs averaging and printing the results
of the measurements (it is a simplified form):
{\small
\begin{verbatim} 
void T_Accum_Measure::Average_Measure_OutResult(){
 AngMomMethod_A->Average_TimeIntObs   ();
 AngMomMethod_A->Out_Result_TimeIntObs(" Set A ");
 AngMomMethod_B->Average_TimeIntObs   ();
 AngMomMethod_B->Out_Result_TimeIntObs(" Set B ");}
\end{verbatim} 
}  

 \section{Class {\tt\bf T\_AngMomMethod}: extraction of 
 physics parameters by using the angular moments method}

  The main data members of the class {\tt T\_AngMomMethod} 
are four pointers to the objects of the class {\tt T\_TimeIntObs} and 
three pointers to the objects of the class {\tt T\_StandardModel\_DG}:
{\small
\begin{verbatim} 
  T_TimeIntObs  *Obs_TG0 ;  //   T , Gp = 0
  T_TimeIntObs  *Obs_TGp ;  //   T , Gp = G'
  T_TimeIntObs  *Obs_T0G0;  //   T0, Gp = 0
  T_TimeIntObs  *Obs_T0Gp;  //   T0, Gp = G'
  T_StandardModel_DG  *SM_DG_TGp_TG0;
  T_StandardModel_DG  *SM_DG_TGp_T0Gp;
  T_StandardModel_DG  *SM_DG_TG0_T0G0;
\end{verbatim} 
}
  Two different definitions of the time-integrated observables
can be considered according to~\cite{SIMUB}: $\tilde{O}$ and $\hat{O}$. 
  The first case corresponds to the zero parameters $\Gamma' = 0$ 
and the second one -  to nonzero $\Gamma'$ defined by the user 
(see the next section about definition of $\Gamma'$). 
  In the program the user can set also two different values 
of the upper time limit $T$ and $T_0$.
 
 Constructor and methods of the class {\tt T\_AngMomMethod} are
{\small
\begin{verbatim} 
  T_AngMomMethod      (...);
  void Init_Angles    (...);
  void Parameter_SIMUB(...)
  void Accumulation_TimeIntObs  ();
  void Average_TimeIntObs       ();
  void Out_Result_TimeIntObs    (const Char_t *cc);
  virtual void Show             (const Char_t *c);
\end{verbatim} 
}
   The constructor calls four constructors for four objects of the 
class {\tt T\_TimeIntObs}.
  Each of the methods listed above calls four methods 
with the same name for the four objects of the class 
{\tt T\_TimeIntObs} and for the three objects of the class 
{\tt T\_StandardModel\_DG}.
  For example,
{\small
\begin{verbatim} 
void T_AngMomMethod::Accumulation_TimeIntObs(){
 Obs_TG0 ->Accumulation_TimeIntObs();
 Obs_TGp ->Accumulation_TimeIntObs();
 Obs_T0G0->Accumulation_TimeIntObs();
 Obs_T0Gp->Accumulation_TimeIntObs();
 SM_DG_TGp_TG0 ->Accumulation();
 SM_DG_TGp_T0Gp->Accumulation();
 SM_DG_TG0_T0G0->Accumulation(); }
\end{verbatim} 
}
  Some features in using the objects  of the classes 
{\tt T\_TimeIntObs} and {\tt T\_StandardModel\_DG} the user can find 
in the method {\tt Parameter\_SIMUB} (this method calls the methods
of {\tt T\_TimeIntObs} objects only), and in the method 
{\tt Average\_TimeIntObs}: 
{\small
\begin{verbatim} 
void T_AngMomMethod::Average_TimeIntObs(){
 Obs_TG0 ->Average_TimeIntObs(); ...;
 SM_DG_TGp_TG0->T_StandardModel_DG_Init(*Obs_TGp, *Obs_TG0 );
 SM_DG_TGp_TG0->Average(); SM_DG_TGp_TG0->Calc_DG(true); ...;}
\end{verbatim} 
}
  Method {\tt T\_StandardModel\_DG\_Init} initiates
the object to extract $\Gamma$  and $\Delta\Gamma$.
  Method {\tt Average} averages values to calculate the 
correlation between two types of the observables.
  Method {\tt Calc\_DG} calculates $\Gamma$  and $\Delta\Gamma$.

 \section{ Class {\tt\bf T\_WeightFunc}: weighting functions}

  Two types of the angular weighting function are proposed
in~\cite{dighe2,SIMUB} to extract the observables:
$w^{(A)}_i(\theta_{l^+},\theta_{K^+},\chi)$ and 
$w^{(B)}_i(\theta_{l^+},\theta_{K^+},\chi)$
($i=\overline{1,6}$).
  As it is shown in~\cite{SIMUB}, it is helpful to consider 
the time dependent weighting functions
which can be written in the general case as 
$$ W^{(A)}_i(\theta_{l^+},\theta_{K^+},\chi, t; \Gamma', T) 
  = \mbox{exp}(\Gamma' t) w^{(A)}_i(\theta_{l^+},\theta_{K^+},\chi)
    \Theta(T-t)
$$ 
and similarly for $W^{(B)}_i$.   
  Weighting function $W^{(A/B)}_i$ is named in the class 
{\tt T\_WeightFunc} as method {\tt wi} ($i = \overline{0,5}$).

  The integer data member {\tt WF\_Type} of the class {\tt T\_WeightFunc} 
sets the type of weighting functions ($1\to A,\, 2\to B$) and 
it is initialized by the class constructor.
  The parameters of the weighting functions $\Gamma'$ and $T$ 
are the constructor input parameters, named 
as {\tt GammaS\_mmDcm1 } (in unit (mm/c)$^{-1}$) 
and {\tt TimeUP\_mmDc} (in unit (mm/c)).  
  The $\Gamma'$ represents a first approximation for the measured 
value $\Gamma$ to be corrected in the current analysis
of the untagged decays.

  Together with method {\tt wi} there are also the methods
with names  {\tt wi\_cTha1}, {\tt wi\_cThb1}, {\tt wi\_Chi} and {\tt wi\_t}
in the class {\tt T\_WeightFunc}.
  These methods represent the derivatives of the function {\tt wi}
with respect to arguments 
$\mbox{cos}\Theta_{\mu^+}$,
$\mbox{cos}\Theta_{K^+}$,
$\chi$ and $t$, correspondently.
 The derivatives are used in class {\tt T\_TimeIntObs} to 
calculate the systematical errors or to estimate 
the detector  response~\cite{SIMUB}.
  
 \section{ Class {\tt\bf T\_TimeIntObs}: six time-integrated observables}

  Class {\tt T\_TimeIntObs} is a derived class to the class 
{\tt T\_WeightFunc}.
  It is used to accumulate and average the time-integrated 
observable, and to extract the physics parameters.

  In the general form by means of the weighting 
function $W^{(X)}_i$ ($X = A,\,B$) on the set of the 
$N(T_{max})$ events we have observables and their statistical errors 
in the following form~\cite{SIMUB}: 
\begin{eqnarray}
\hat{b}_i^{(exp)}(\Gamma',\,T) = \frac{1}{N(T_{max})}\sum_{j=1}^{N(T)}\, 
                       W^{(X)}_{ij}(\Gamma',\, T),
\label{observ_hat_exp} \\
\delta\hat{b}_i^{(exp)}(\Gamma',\,T) = \frac{1}{N(T_{max})}
\sqrt{\sum_{j=1}^{N(T)}[\hat{b}_i^{(exp)}(\Gamma',\,T) - W^{(X)}_{ij}(\Gamma',\, T)]^2},
\nonumber
\end{eqnarray}
where 
$W^{(X)}_{ij}(\Gamma',T)\equiv W^{(X)}_i(\theta^j_{l^+},\theta^j_{K^+},\chi^j, t^j;\Gamma',T)$ 
is a value of the weighting function for the $j$-th event. 
  Including the time dependent  $\Theta(T-t)$-function 
in the weighting function allows to keep the 
time informative contents in the value
$\hat{b}_i^{(exp)}(\Gamma',\,T)$ and in the same time
to have good statistical errors which are caused by 
a large size of the proper time interval $[0,\,T]$.

  The class constructor {\tt T\_TimeIntObs} is defined by the 
following signature:
{\small
\begin{verbatim} 
T_TimeIntObs(char *cc, int Type_WF, 
             double GammaS_mmDcm1, TimeUP_mmDc, TimeMAX_mmDc,
             bool   DEBUG, bRealData, T_Physics *P_hys);
\end{verbatim} 
}
  Three parameters {\tt Type\_WF, GammaS\_mmDcm1, TimeUP\_mmDc}
are sent to the mother class {\tt T\_WeightFunc} 
(see the previous section).
  Parameter {\tt TimeMAX\_mmDc} defines the value $T_{max}$.

  By means of the logical parameter {\tt bRealData} the user can set 
either the mode for the real data or the mode for the generator data.
  The latest case does not use the method parameter {\tt P\_hys} and data member
{\tt Phys} of the class will be filled by the generator data 
taken from the first event by the method {\tt Parameter\_SIMUB}.

  In case of {\tt bRealData = true} the user has to define 
the object of the class {\tt T\_Physics} in the main program and send 
it into the object of the class {\tt T\_Accum\_Measure} defined 
in the main program.

  The constructor and list of the methods of the class 
are as follows:
{\small
\begin{verbatim} 
  T_TimeIntObs(char *cc, int  Type_WF,
               double GammaS_mmDcm1, TimeUP_mmDc, TimeMAX_mmDc,
               bool DEBUG, b_RealData, T_Physics *P_hys);
  void FinalObserv_TimeIntObs (const Char_t *c);
  void Parameter_SIMUB(double delta1, delta2, TotalPhaseWeak,
    nA0_02, nA0_P2, nA0_T2, Gamma_Bmes_GeV, Gp_GeV, Gm_GeV, Delta_G_GeV, 
    Gen_dcTb1, Gen_dcTa1, Gen_dChi, Gen_dt);
  void Accumulation_TimeIntObs();
  void Average_TimeIntObs     ();
  void Out_Result_TimeIntObs  (const Char_t *c);
  virtual void Show           (const Char_t *c);
\end{verbatim} 
}

  If we compare this list with  the list of the methods of 
the class {\tt T\_AngMomMethod}, it is possible to note
that the last 5 methods have the same names as the method names 
of the class {\tt T\_AngMomMethod} because the calls of these methods 
are the main purpose of the corresponding methods 
of the class {\tt T\_AngMomMethod}
(see explanation in the description of the class  {\tt T\_AngMomMethod}).

  Method {\tt FinalObserv\_TimeIntObs} is called by method
{\tt  Out\_Result\_TimeIntObs} and calculates the final observables
or their combinations according to the formulae~\cite{SIMUB}:
\begin{eqnarray}
&& |A_0(0)|^2 = \frac{\hat{b}^{(exp)}_1(\Gamma',T)}
                {\hat{b}^{(exp)}_1(\Gamma',T)+\hat{b}^{(exp)}_2(\Gamma',T)
                +\hat{b}^{(exp)}_3(\Gamma',T)/\hat{\gamma}}\,,
\nonumber \\
&& |A_{||}(0)|^2 = \frac{\hat{b}^{(exp)}_2(\Gamma',\,T)}
                   {\hat{b}^{(exp)}_1(\Gamma',T)+\hat{b}^{(exp)}_2(\Gamma',T)
                   +\hat{b}^{(exp)}_3(\Gamma',T)/\hat{\gamma}}\,,
\nonumber \\
&& |A_{\bot}(0)|^2 = \frac{\hat{b}^{(exp)}_3(\Gamma',T)/\hat{\gamma}}
                     {\hat{b}^{(exp)}_1(\Gamma',T)+\hat{b}^{(exp)}_2(\Gamma',T)
                     +\hat{b}^{(exp)}_3(\Gamma',T)/\hat{\gamma}}\,,
\nonumber \\
&&  \mbox{cos}(\delta_2 - \delta_1) = 
    \frac{\hat{b}^{(exp)}_5(\Gamma',T)}
    {\sqrt{\hat{b}^{(exp)}_1(\Gamma',T)\,\hat{b}^{(exp)}_2(\Gamma',T)}}\,,
\label{observ_tild_ut_3}
\end{eqnarray}
where we consider the initial amplitudes normalized as 
$|A_0(0)|^2+|A_{||}(0)|^2+|A_\bot (0)|^2$ $ = 1$.
   We have introduced the function $\gamma$:
\begin{eqnarray}
 && \hat{\gamma}(\Delta\Gamma_L,\Delta\Gamma_H,\mbox{cos}\phi_c^{(s)},T) 
       = \frac{\hat{G}_H}{\hat{G}_L}\,,
\label{gamma} \\
&& \hat{G}_{\normalsize\frac{L}{H}} = (1\pm\mbox{cos}\phi_c^{(s)})\,
     \frac{\mbox{e}^{ \Delta\Gamma_LT/2}-1}{\Delta\Gamma_L}
    -(1\mp\mbox{cos}\phi_c^{(s)})\,
     \frac{\mbox{e}^{-\Delta\Gamma_HT/2}-1}{\Delta\Gamma_H}\,.
\label{G_LH}
\end{eqnarray}
$\Delta\Gamma_{L/H}$ are measured parameters related with physics 
values of widths of light and heavy states $\Gamma_{L/H}$  by
\begin{equation}
\Delta\Gamma_L = 2(\Gamma'  -\Gamma_L)\,,\qquad
\Delta\Gamma_H =-2(\Gamma'  -\Gamma_H)\,.
\label{Delta_G_L/H}
\end{equation} 

  The method {\tt FinalObserv\_TimeIntObs} also calculates
two values:
\begin{eqnarray}
 \mbox{sin}\phi_c^{(s)}\,\mbox{cos}\delta_{1,2} =
          \frac{\hat{b}^{(exp)}_{4,6}(\Gamma',T)}
               {\sqrt{\hat{b}^{(exp)}_{2,1}(\Gamma',T)\hat{b}^{(exp)}_3(\Gamma',T)}}\,
	       \hat{\beta}.
\label{sphi}
\end{eqnarray}
 where
\begin{eqnarray}
 \hat{\beta}(\Delta\Gamma_L,\Delta\Gamma_H,\mbox{cos}\phi_c^{(s)},T) 
  =  \frac{\sqrt{\hat{G}_L\,\hat{G}_H}}{\hat{Z}}\,,\quad
 \hat{Z}  = \frac{1-\mbox{e}^{ \Delta\Gamma_LT/2}}{\Delta\Gamma_L}
      +\frac{1-\mbox{e}^{-\Delta\Gamma_HT/2}}{\Delta\Gamma_H}\,.
\label{beta}
\end{eqnarray}
  Eqs.~(\ref{sphi}) contain the weak phase in the left- and right-handed 
side and might be helpful in case of large values of weak phase $\phi_c^{(s)}$
when a large violation of the Standard Model predictions,
$\phi_c^{(s)} \approx 0.03$, takes place.

  Values $\hat{\gamma}$ and $\hat{\beta}$ are calculated 
in the data member {\tt T\_Physics *Phys} of the class {\tt T\_TimeIntObs}.

  The results of extraction of final observables~(\ref{observ_tild_ut_3}) 
and ~(\ref{sphi}) are given in Tabls.~1 and ~2. 

   The direct numerical calculations have shown that the difference between 
the values of observables $\hat{b}_i(T)$ $(i=1,2,3,5)$, calculated with 
$\phi^{(s)}_c=0$ and $\phi^{(s)}_c=0.04$, does not exceed 0.01\%.
   Even in case of statistics of 100~000 events this difference is negligibly
small as compared with statistical errors for these observables.
   Therefore, the assumption $\phi^{(s)}_c\approx 0$ is a good approximation
in case of Standard Model. 
   
   To determine the initial transversity amplitudes in case of SM,  
it is convenient to use $\Gamma' = 0$ and for large $T$ we have
$ \hat{\gamma} \approx \Gamma_L/\Gamma_H$.

\begin{table}
\begin{center}
\caption{Time-integrated observables 
     obtained by data member {\tt Obs\_TG0} of class  {\tt T\_AngMomMethod}.
     The sample of 300\,000 (3\,000 in parentheses) 
     untagged $B_s^0\to J/\psi\phi$ events from SIMUB generator
     with the "main setting" of physics parameters (see section 5) is used}
\label{tab:Obs_TG0_1}
\vspace*{5mm}
\begin{tabular}{|c|c|c|c|c|}
\hline
$i$ &  $\hat{b}_i(0,T)_{T=T_{max}\to 0}$,  & $\hat{b}_i(0,T)_{T=T_{max}=2}$, & $\hat{b}_i^{(exp)}(0,\,T)_{T=T_{max}=2}$,& Statistical \\ 
    &   Eq.~(\ref{observ_hat})         &  Eq.~(\ref{observ_hat}) &  Eq.~(\ref{observ_hat_exp}) & errors \\ 
\hline
1   & 0.54   & 0.5225 & 0.5207 (0.488) & 0.0014 (0.014)   \\
2   & 0.30   & 0.2903 & 0.2940 (0.380) & 0.0021 (0.021)   \\
3   & 0.16   & 0.1873 & 0.1849 (0.132) & 0.0020 (0.020)   \\
4   & 0.0088 & 0.0009 & 0.0032 (0.011) & 0.0020 (0.020)   \\
5   & -0.4025&-0.3894 &-0.3916 (-0.316) & 0.0029 (0.029)  \\
6   & -0.0118&-0.0012 &-0.0021 (-0.002) & 0.0031 (0.031)  \\
\hline
\end{tabular}
\end{center}
\end{table}
  The first observable in the list of four observables of the 
class  {\tt T\_AngMomMethod} is {\tt Obs\_TG0} which is
initialized with setting  $\Gamma' = 0$ and with
$T=T_{max}$  as the upper time limit.
  The results from  {\tt Obs\_TG0} are presented by two tables
in the listing (see section "Test Run Input and Output").

  These data are given in Tabls.~\ref{tab:Obs_TG0_1} and ~\ref{tab:Obs_TG0_2}.
\begin{table}
\begin{center}
\caption{Determination of values~(\ref{observ_tild_ut_3}) 
	 and~(\ref{sphi})
         by using the observables 
         $\hat{b}^{(exp)}_i(0,T)$ ($T = 2/1/0.5/0.25/0.125$ mm/c$^{-1}$)
	 for 300\,000  $B_s^0\to J/\psi\phi$ 
	 events with the "main setting"  (see section 5) 
	 of physics parameters}
\label{tab:Obs_TG0_2}
\vspace*{5mm}
\begin{tabular}{|c|c|c|c|c|}
\hline
Parameter                               &Input & Extracted value & Statistical  \\ 
                                        &      & by~(\ref{observ_tild_ut_3}),~(\ref{sphi})  & error \\ 
\hline
$A_{||}^2/A_0^2  $                      &  0.556& 0.565/0.568/0.565/0.568   & 0.005/0.005/0.006/0.007 \\
$A_{\bot}^2/A_0^2$                      &  0.296& 0.294/0.293/0.293/0.288   & 0.004/0.004/0.005/0.006 \\
$|A_0(0)|^2$                            &   0.54& 0.538/0.538/0.538/0.539   & 0.001/0.001/0.002/0.002\\
$|A_{||}|^2$                            &   0.30& 0.304/0.305/0.304/0.306   & 0.002/0.002/0.003/0.003 \\
$|A_\bot|^2$                            &   0.16& 0.158/0.157/0.158/0.155   & 0.002/0.002/0.002/0.003 \\
$\mbox{cos}(\delta_2-\delta_1)$         &    -1 &-1,001/-1,000/-0.998/-0.999& 0.009/0.009/0.010/0.012  \\
$\mbox{cos}\delta_1\mbox{sin}\phi_c^{(s)}$& 0.04&  0.15/0.17/0.16/-0.04     & 0.09/0.12/0.22/0.50 \\
$\mbox{cos}\delta_2\mbox{sin}\phi_c^{(s)}$&-0.04& -0.07/-0.07/0.03/-0.03    & 0.10 /0.14/0.26/0.61  \\
\hline
\end{tabular}
\end{center}
\end{table}
  
  As it is shown in the first two columns of Tabl.~\ref{tab:Obs_TG0_1},
the time dependence of observables is essential in case the number of events 
is more than $3\,000$ because for the $3\,000$ events the differences 
$\hat{b}_i(0,T)_{T=T_{max}\to 0} - \hat{b}_i(0,T)_{T=T_{max} = 2}$
are comparable with statistical errors.
 This conclusion depends on the value of width difference $\Delta\Gamma$.
 In case of the $B_d^0\to J/\psi K^*$ channel the Standard Model predicts  
a small value of $\Delta\Gamma$ ("main setting" for 
$B_d^0$ $\Delta\Gamma/\Gamma = -0.01$, see  section 5) 
and in this case 
$\hat{b}_i(0,T)_{T=T_{max}\to 0} \approx \hat{b}_i(0,T)_{T=T_{max}= 2}$.

  The second observable in class {\tt T\_AngMomMethod} is {\tt Obs\_TGp} 
which is considered here with parameter $\Gamma' = \Gamma$ 
($\Gamma$ is a true value of the B meson width)
and with $T_{max}$  as the upper lifetime limit.
  The results from listing tables of observable {\tt Obs\_TGp} 
are shown here in Tabl.~\ref{tab:Obs_TGp_1} and ~\ref{tab:Obs_TGp_2}
and in listing of section "Test Run Input and Output".
\begin{table}
\begin{center}
\caption{Time-integrated observables 
     obtained by data member {\tt Obs\_TGp} of class  {\tt T\_AngMomMethod}
     for 300\,000 (3\,000 in  parentheses)
     untagged $B_s^0\to J/\psi\phi$ events with
     the "main setting" (see section 5) of physics parameters}
\label{tab:Obs_TGp_1}
\vspace*{5mm}
\begin{tabular}{|c|c|c|c|c|}
\hline
$i$   & $\hat{b}_i(\Gamma,T_{max})$, Eq.~(\ref{observ_hat}) & $\hat{b}_i^{(exp)}(\Gamma,\,T_{max})$, Eq.~(\ref{observ_hat_exp}) & Stat.err. \\ 
\hline
1      & 2.117 & 2.139 (1,95) & 0.014  (0.12)     \\
2      & 1.176 & 1.157 (1,40) & 0.019  (0.17)     \\
3      & 0.989 & 1.024 (0.78) & 0.020  (0.16)    \\
4      & 0.010 & 0.035 (0.07) & 0.018  (0.19)     \\
5      &-1.578 &-1.603 (-0.94) & 0.028 (0.24)      \\
6      &-0.013 &-0.010 (0.11) & 0.028  (0.31)    \\
\hline
\end{tabular}
\end{center}
\end{table}
\begin{table}
\begin{center}
\caption{Determination of values
          ~(\ref{observ_tild_ut_3}) and~(\ref{sphi})
         by using the observables 
         $\hat{b}^{(exp)}_i(\Gamma,T_{max})$ for 300\,000  $B_s^0\to J/\psi\phi$ 
	 events with the  "main setting" (see section 5) 
	 of physics parameters}
\label{tab:Obs_TGp_2}
\vspace*{5mm}
\begin{tabular}{|c|c|c|c|c|}
\hline
Parameter                               &Input value& Extracted value & Stat.error \\ 
\hline
$A_{||}^2/A_0^2  $                      &  0.556& 0.5409 &  0.010  \\
$A_{\bot}^2/A_0^2$                      &  0.296& 0.3038 &  0.0066 \\
$|A_0(0)|^2$                            &   0.54& 0.5421 &  0.0030 \\
$|A_{||}|^2$                            &   0.30& 0.2932 &  0.0045 \\
$|A_\bot|^2$                            &   0.16& 0.1647 &  0.0034  \\
$\mbox{cos}(\delta_2-\delta_1)$         &    -1 & -1,019&  0.020  \\
$\mbox{cos}\delta_1\mbox{sin}\phi_c^{(s)}$& 0.04& 0.138 &  0.073 \\
$\mbox{cos}\delta_2\mbox{sin}\phi_c^{(s)}$&-0.04& -0.029&  0.081  \\
\hline
\end{tabular}
\end{center}
\end{table}
  It is better to use observables  $\hat{b}^{(exp)}_i(0,T_{max})$
to extract the initial observables ~(\ref{observ_tild_ut_3}) and~(\ref{sphi}) 
(compare the statistical errors in Tabl.~\ref{tab:Obs_TG0_2} 
and~\ref{tab:Obs_TGp_2}).

   Extraction of combinations 
$\mbox{cos}\delta_{1,2}\mbox{sin}\phi_c^{(s)}$ is shown in Tabl.~\ref{tab:Phi}.

\begin{table}
\begin{center}
\caption{Determination of values depended on weak phase by using the
    observables $\hat{b}^{(exp)}_i(\Gamma,T_{max})$ extracted from 
    Monte Carlo data for two values of $\phi_c^{(s)} =$ 0.4 and  0.04.
    The events sample has been generated for the case of "main setting"
    for physics parameters}
\label{tab:Phi}
\vspace*{5mm}
\begin{tabular}{|c|c|c|c|c|}
\hline
Parameter                               &Input value&100\,000 events &200\,000 events &400\,000 events\\ 
\hline
$\mbox{cos}\delta_1\mbox{sin}\phi_c^{(s)}$& 0.04&$ 0.24\pm 0.14$&$0.13\pm 0.09$&$ 0.12\pm 0.07$ \\
$\mbox{cos}\delta_2\mbox{sin}\phi_c^{(s)}$&-0.04&$-0.05\pm 0.15$&$0.10\pm 0.11$&$-0.03\pm 0.07$ \\
\hline
$\mbox{cos}\delta_1\mbox{sin}\phi_c^{(s)}$& 0.389&$ 0.49\pm 0.16$&$ 0.44\pm 0.11$&$ 0.35\pm 0.08 $ \\
$\mbox{cos}\delta_2\mbox{sin}\phi_c^{(s)}$&-0.389&$-0.29\pm 0.16$&$-0.31\pm 0.12$&$-0.41\pm 0.09 $ \\
\hline
\end{tabular}
\end{center}
\end{table}

 \section{ Class {\tt\bf T\_Physics}: physics values}

   For observables $\hat{b}_i(\Gamma',\,T)$ we have the following 
formulae~\cite{SIMUB}:
\begin{eqnarray}
       &&
\hat{b}_1(\Gamma',\,T)=|A_0(0)|^2\,\hat{G}_L(\Gamma',\,T)/\tilde{L}(T_{max})\,,
\nonumber \\ && 
\hat{b}_2(\Gamma',\,T)=|A_{||}(0)|^2\,\hat{G}_L(\Gamma',\,T)/\tilde{L}(T_{max})\,,
\nonumber \\ && 
\hat{b}_3(\Gamma',\,T)=|A_{\bot}(0)|^2\,\hat{G}_H(\Gamma',\,T)/\tilde{L}(T_{max})\,,
\nonumber \\ && 
\hat{b}_4(\Gamma',\,T)=|A_{||}(0)|\,|A_{\bot}(0)|\,\hat{Z}(\Gamma',\,T)\,
                          \mbox{cos}\delta_1\,\mbox{sin}\phi_c^{(s)}/\tilde{L}(T_{max})\,,
\nonumber \\ && 
\hat{b}_5(\Gamma',\,T)=|A_0(0)|\,|A_{||}(0)|\,\hat{G}_L(\Gamma',\,T)\,
                          \mbox{cos}(\delta_2-\delta_1)/\tilde{L}(T_{max})\,, 
\nonumber \\ && 
\hat{b}_6(\Gamma',\,T)=|A_0(0)|\,|A_{\bot}(0)|\,\hat{Z}(\Gamma',\,T)\,
                         \mbox{cos}\delta_2\,\mbox{sin}\phi_c^{(s)}/\tilde{L}(T_{max})\,,
\label{observ_hat}
\end{eqnarray}
where  $\tilde{L}(T_{max})$ is a normalization factor, which has the form:
\begin{eqnarray}
\tilde{L}(T_{max}) &=&
 (|A_0(0)|^2 + |A_{||}(0)|^2)\,\hat{G}_L(0,\,T_{max}) + |A_\bot(0)|^2\,\hat{G}_H(0,T_{max})\,,
\label{norm-fact}
\end{eqnarray}
  From expresions~(\ref{observ_hat}) one can see 
that  $\hat{b}_1(\Gamma',\,T) + \hat{b}_2(\Gamma',\,T)+ \hat{b}_3(\Gamma',\,T) = 1$.

  In the general case with sizable weak phase $\phi_c^{(s)}$
to extract the physics values we can use the following equations:
\begin{eqnarray}
\frac{\hat{b}^{(exp)}_{1,2,5}(\Gamma', T)}{\hat{b}^{(exp)}_{1,2,5}(\Gamma'', T_0)}
 &=&  \mu_{L}(\Delta\Gamma_L, \Delta\Gamma_H, \mbox{cos}\phi_c^{(s)};[\Gamma', T],[\Gamma'', T_0]);
\label{DG_mu_equa_125}
\\
\frac{\hat{b}^{(exp)}_3(\Gamma',T)}{\hat{b}^{(exp)}_3(\Gamma'',T_0)} 
 &=& \mu_{H}(\Delta\Gamma_L, \Delta\Gamma_H, \mbox{cos}\phi_c^{(s)};[\Gamma', T],[\Gamma'', T_0])
\label{DG_mu_equa_3}
\\
\frac{\hat{b}^{(exp)}_{4,6}(\Gamma',T)}{\hat{b}^{(exp)}_{4,6}(\Gamma'',T_0)} 
 &=& \rho(\Delta\Gamma_L, \Delta\Gamma_H; [\Gamma', T],[\Gamma'', T_0]),
\label{DG_mu_equa_46}
\end{eqnarray}
where 
\begin{eqnarray}
 \mu_{L/H}  &\equiv&
  \frac{\hat{G}_{L/H}(\Delta\Gamma_L, \Delta\Gamma_H, \mbox{cos}\phi_c^{(s)};[\Gamma', T])}
   {\hat{G}_{L/H}(\Delta\Gamma_L, \Delta\Gamma_H, \mbox{cos}\phi_c^{(s)};[\Gamma'', T_0])},
\label{mu}\\
 \rho  &\equiv&
  \frac{\hat{Z}(\Delta\Gamma_L, \Delta\Gamma_H;[\Gamma', T])}
   {\hat{Z}(\Delta\Gamma_L, \Delta\Gamma_H;[\Gamma'', T_0])},
\label{mu}
\end{eqnarray}
  Therefore, to extract tree parameters 
$\Delta\Gamma_L$, $\Delta\Gamma_H$  and
$\mbox{cos}\phi_c^{(s)}$ we need 
the experimental values of the time-integrated 
observables $\hat{b}^{(exp)}_i$ only.

  In case of small values $\hat{b}^{(exp)}_{4,6}$ we have a small 
$\mbox{sin}\phi_c^{(s)}\,\mbox{cos}\delta_{1,2}$, but it does not
mean the small weak phase $\phi_c^{(s)}$ as it is predicted 
by the Standard Model.
  The opposite case of the sizable observables $\hat{b}^{(exp)}_{4,6}$
will directly show the signal of the beyond Standard Model physics.
  In the last case we can extract $\Delta\Gamma_L$ and $\Delta\Gamma_H$
from the two equations~(\ref{DG_mu_equa_46}).
  If we use $\Gamma'$ as the true value of the B meson width 
(found from other sources), then we will have a single unknown value
$\Delta\Gamma_L = \Delta\Gamma_H \equiv \Delta\Gamma$.
  Simplification of the system of equations
~(\ref{DG_mu_equa_125})-~(\ref{DG_mu_equa_46})
in case of the Standard Model will be considered below.

  After extraction of $\Delta\Gamma_L$, $\Delta\Gamma_H$  and
$\mbox{cos}\phi_c^{(s)}$ we can calculate the 
values $\hat{\gamma}$~(\ref{gamma}) and $\hat{\beta}$~(\ref{beta}).
  The values  $\hat{\gamma}$ and $\hat{\beta}$ are calculated
in class {\tt T\_Physics} and 
used in the method {\tt FinalObserv\_TimeIntObs} 
of the class {\tt T\_TimeIntObs} to calculate 
the final observables~(\ref{observ_tild_ut_3})  and ~(\ref{sphi}).
  It is the first goal of the class {\tt T\_Physics}.

  The second goal of the class {\tt T\_Physics} is to calculate
exact theoretical values of the observables~(\ref{observ_hat})
to compare them with the approximate values obtained 
by~(\ref{observ_hat_exp}) in the class {\tt T\_TimeIntObs}.

  For these purposes the class  {\tt T\_Physics} collects 
all the physics values including transversity amplitudes 
with its errors and three user parameters 
$\Gamma'$, $T$ and $T_{max}$.

  In the main method {\tt Init} of the class {\tt T\_Physics} 
one can see the scenario to calculate of the physics values
described above.
  The list of the input parameters of the method {\tt Init} includes
8 physics parameters:
{\tt delta1}  and {\tt delta2} (strong CP-conserving phase 
$\delta_{1,2}$),
{\tt TotalPhaseWeak} (weak CP-violating phase $\phi_c^{(s)}$),  
{\tt nA0\_02},  {\tt nA0\_P2} ,  {\tt nA0\_T2} 
(initial transversity amplitudes squared 
$|A_0(0)|^2$, $|A_{||}(0)|^2$ 
and $|A_{\bot}(0)|^2 = 1 - |A_0(0)|^2 - |A_{||}(0)|^2$),
{\tt Gp} and {\tt Gm} 
(widths of light and heavy B meson states $\Gamma_L$ 
and $\Gamma_H$), their errors, and 3 user defined parameters: 
{\tt TimeUP} (upper time limit $T$), {\tt  TimeMAX} 
(maximal time $T_{max}$) and {\tt Gprime} ($\Gamma'$)
to use them for calculation by formulae~(\ref{observ_hat}).
  Data member {\tt oEk\_hat} ($k=\overline{0,5}$) 
of the class {\tt T\_Physics} are the observables~(\ref{observ_hat}) 
in case of $\Gamma'$  defined by user while {\tt oEk\_tilde} 
($k=\overline{0,5}$) are the observables~(\ref{observ_hat}) 
in case of $\Gamma'=0$.

 \section{ Class {\tt\bf T\_StandardModel\_DG }:
   extraction of $\Gamma$  and $\Delta\Gamma$
   in case of the Standard Model}

  The name of the class {\tt T\_StandardModel\_DG} means that
it includes the extraction of $\Gamma$  and $\Delta\Gamma$ 
in case of the Standard Model expectation for the weak phase:
$\phi_c^{(s)} \approx 0$.
  In this case we may not use the small values of the observables 
$\hat{b}^{(exp)}_{4,6}$ and, 
according to~(\ref{G_LH}) and~(\ref{beta}), we have:
$\mu_L = \mu_L(\Delta\Gamma_L;[\Gamma', T],[\Gamma'', T_0])$,
$\mu_H = \mu_H(\Delta\Gamma_H;[\Gamma', T],[\Gamma'', T_0])$.
  It means that extraction of $\Delta\Gamma_L$ and  $\Delta\Gamma_H$
can be performed by numerical solving of the separate equations.


  Three cases numbered as {\tt KK = 1,2,3} 
(see method {\tt mu\_DG\_Equa\_Solut})
are realized in the class {\tt T\_StandardModel\_DG}:
{\tt KK = 1} in case of $\Gamma' \ne 0$, $\Gamma''  = 0$, $T_0 =T$
(used in data member {\tt SM\_DG\_TGp\_TG0} 
of the class {\tt T\_AngMomMethod}), 
{\tt KK = 2} in case of $\Gamma'   =  \Gamma''  = 0$, $T_0 \ne T$
(used in data member {\tt SM\_DG\_TG0\_T0G0}), 
 and {\tt KK = 3} in case of $\Gamma' = \Gamma'' \ne 0$, $T_0 \ne T$
(used in {\tt SM\_DG\_TGp\_T0Gp}).
  Only the case {\tt KK = 1} is presented
in listing which is shown in the section "Test Run Input and Output".
  In each case we have 3 equations~(\ref{DG_mu_equa_125}) 
to determine $\Delta\Gamma_L$ 
(names {\tt DGL\_0}, {\tt DGL\_1} and {\tt DGL\_4}
in output listing)
and one equation~(\ref{DG_mu_equa_3}) 
for  $\Delta\Gamma_H$ (with name {\tt DGH\_2} in output listing).

  Combining three values  $\Delta\Gamma_L$ from ~(\ref{DG_mu_equa_125}) 
with one value $\Delta\Gamma_H$ from ~(\ref{DG_mu_equa_3}),
we have three value $\Delta\Gamma = 0.5(\Delta\Gamma_L+\Delta\Gamma_H)$
(names {\tt DG\_0}, {\tt DG\_1} and {\tt DG\_4} in listing of section
"Test Run Input and Output" )
and three value $\Gamma = \Gamma' - 0.25(\Delta\Gamma_L-\Delta\Gamma_H)$
(names {\tt G\_0}, {\tt G\_1} and {\tt G\_4} in listing)
corresponding to combinations of indices [1,3], [2,3] and [5,3] of observables.

  Three cases {\tt KK=1,2,3} of extraction of 
$\Gamma$ and $\Delta\Gamma$ are presented  as three tables in listing.

  The results of the first method of extraction of 
$\Gamma$ and $\Delta\Gamma$ ({\tt KK=1}) weakly dependend on user
defined $\Gamma'$ value
(see Tabl.~\ref{tab:Gamma_KK_1} obtained by {\tt SM\_DG\_TGp\_TG0}).
  Due to weak dependences of the result 
on $\Gamma'$ we fix this parameter in the 
following tables as $\Gamma' = \Gamma$.
\begin{table}
\begin{center}
\caption{Determination of values $\Gamma$ and $\Delta\Gamma$
         by using the object {\tt SM\_DG\_TGp\_TG0} ({\tt KK = 1})
         for 300\,000   $B_s^0\to J/\psi\phi$ 
	 events with the  "main setting" of physics parameters.
	 Combinations of indices [1,3] [2,3] [5,3] are explained 
	 in section 11}
\label{tab:Gamma_KK_1}
\vspace*{5mm}
\begin{tabular}{|c|c|c|c|c|c|}
\hline
Parameter                    & $\Gamma'$& Input    & [1,3]            & [2,3]           & [5,3] \\ 
\hline
$\Gamma$, [mm/c]$^{-1}$      & 2.278    &  2.278   &$ 2.228\pm 0.015$ &$ 2.268\pm 0.020$&$ 2.231\pm 0.020$ \\
$\Delta\Gamma$, [mm/c]$^{-1}$& 2.278    & -0.456   &$-0.506\pm 0.030$ &$-0.587\pm 0.040$&$-0.512\pm 0.040$  \\
\hline
$\Gamma$, [mm/c]$^{-1}$      & 2.392    &  2.278   &$2.227\pm 0.015$ &$ 2.268\pm 0.020$&$2.230\pm 0.020$ \\
$\Delta\Gamma$, [mm/c]$^{-1}$& 2.392    & -0.456   &$-0.508\pm 0.030$ &$-0.591\pm 0.040$&$-0.515\pm 0.041$  \\
\hline
$\Gamma$, [mm/c]$^{-1}$      & 2.164    &  2.278   & $2.229\pm 0.015$ & $2.268\pm 0.019$ & $2.232\pm 0.020$ \\
$\Delta\Gamma$, [mm/c]$^{-1}$& 2.164    & -0.456   & $-0.504\pm 0.029$ & $-0.582\pm 0.039$ & $-0.509\pm 0.039$  \\
\hline
\end{tabular}
\end{center}
\end{table}

  The results of the second method of extraction of 
$\Gamma$ and $\Delta\Gamma$ ({\tt KK = 2}) depend on the user
defined time up limits $T$ and $T_0$
(see Tabl.~\ref{tab:Gamma_KK_2} obtained by {\tt SM\_DG\_TG0\_T0G0}).
  The best choice is $T_0 = 0.5$ mm/c in case of maximal $T = T_{max} = 2$ mm/c.
\begin{table}
\begin{center}
\caption{Determination of values $\Gamma$ and $\Delta\Gamma$
         by using the object {\tt SM\_DG\_TG0\_T0G0} ({\tt KK = 2})
         for 300\,000   $B_s^0\to J/\psi\phi$ 
	 events with the  "main setting" of physics parameters.
	 Combinations of indices [1,3] [2,3] [5,3] are explained 
	 in section 11}
\label{tab:Gamma_KK_2}
\vspace*{5mm}
\begin{tabular}{|c|c|c|c|c|c|}
\hline
Parameter                    & $T_0; T$,   & Input    & [1,3]            & [2,3]           & [5,3] \\ 
                & $\frac{\mbox{mm}}{c}$  & $[\frac{\mbox{mm}}{c}]^{-1}$  &   $[\frac{\mbox{mm}}{c}]^{-1}$     &  $[\frac{\mbox{mm}}{c}]^{-1}$    &  $[\frac{\mbox{mm}}{c}]^{-1}$  \\ 
\hline
$\Gamma$       &  1; 2    &  2.278   &$2.246\pm 0.017$ &$2.281\pm 0.022$&$2.255\pm 0.022$ \\
$\Delta\Gamma$ &  1; 2    & -0.456   &$-0.478\pm 0.033$ &$-0.549\pm 0.043$&$-0.498\pm 0.043$  \\
\hline
$\Gamma$       &  0.5; 2  &  2.278   & $2.242\pm 0.015$ & $2.272\pm 0.019$ & $2.243\pm 0.019$ \\
$\Delta\Gamma$ &  0.5; 2  & -0.456   & $-0.485\pm 0.029$ & $-0.544\pm 0.038$ & $-0.487\pm 0.039$ \\
\hline
$\Gamma$       &  0.25; 2&  2.278   & $ 2.240\pm 0.021$ & $ 2.249\pm 0.025$ & $2.242\pm 0.025$ \\
$\Delta\Gamma$ &  0.25; 2& -0.456   & $-0.509\pm 0.042$ & $-0.527\pm 0.049$ & $-0.512\pm 0.050$  \\
\hline
$\Gamma$       &  0.125; 2 &  2.278   & $2.257\pm 0.028$ & $2.236\pm 0.032$ & $2.210\pm 0.033$ \\
$\Delta\Gamma$ &  0.125; 2 & -0.456   & $-0.477\pm 0.056$ & $-0.434\pm 0.065$ & $-0.383\pm 0.066$  \\
\hline
\end{tabular}
\end{center}
\end{table}

  The results of the third method of extraction of 
$\Gamma$ and $\Delta\Gamma$ ({\tt KK = 3}) depend 
on time upper limits $T$ and $T_0$ defined by the user
(see Tabl.~\ref{tab:Gamma_KK_3} obtained by {\tt SM\_DG\_TGp\_T0Gp}).
  The best choice is $T_0 = 0.25$ mm/c in case of maximal $T = T_{max} = 2$ mm/c.
  The last method of extraction of $\Gamma$ and $\Delta\Gamma$ is better
among all the considered methods.

\begin{table}
\begin{center}
\caption{Determination of values $\Gamma$ and $\Delta\Gamma$
         by using the object {\tt SM\_DG\_TGp\_T0Gp} ({\tt KK = 3})
         for 300\,000   $B_s^0\to J/\psi\phi$ 
	 events with the  "main setting" of physics parameters.
	 Combinations of indices [1,3] [2,3] [5,3] are explained 
	 in section 11}
\label{tab:Gamma_KK_3}
\vspace*{5mm}
\begin{tabular}{|c|c|c|c|c|c|}
\hline
Parameter                    & $T_0; T$,   & Input    & [1,3]            & [2,3]           & [5,3] \\ 
                & $\frac{\mbox{mm}}{c}$  & $[\frac{\mbox{mm}}{c}]^{-1}$  &   $[\frac{\mbox{mm}}{c}]^{-1}$     &  $[\frac{\mbox{mm}}{c}]^{-1}$    &  $[\frac{\mbox{mm}}{c}]^{-1}$  \\ 
\hline
$\Gamma$       &  1; 2    &  2.278   &$ 2.225\pm 0.019 $ &$2.279\pm 0.026 $&$2.238\pm 0.026 $ \\
$\Delta\Gamma$ &  1; 2    & -0.456   &$-0.514\pm 0.039 $ &$-0.622\pm 0.053 $&$ -0.541\pm 0.053 $  \\
\hline
$\Gamma$       &  0.5; 2  &  2.278   & $ 2.228\pm 0.015$ & $ 2.268\pm 0.020 $ & $ 2.231\pm 0.020 $ \\
$\Delta\Gamma$ &  0.5; 2  & -0.456   & $-0.506\pm 0.030$ & $-0.587\pm 0.040 $ & $-0.512\pm 0.040 $ \\
\hline
$\Gamma$       &  0.25; 2 &  2.278   & $2.231 \pm 0.014 $ & $2.263\pm 0.018 $ & $2.237\pm  0.018 $ \\
$\Delta\Gamma$ &  0.25; 2 & -0.456   & $-0.508\pm 0.029 $ & $-0.573\pm 0.036 $ & $-0.520\pm 0.037 $  \\
\hline
$\Gamma$       &  0.125; 2&  2.278   & $2.238\pm 0.016$ & $2.256\pm 0.019$ & $2.222\pm 0.020$ \\
$\Delta\Gamma$ &  0.125; 2& -0.456   & $-0.496\pm 0.032$ & $-0.532\pm 0.039$ & $-0.465\pm 0.039$  \\
\hline
\end{tabular}
\end{center}
\end{table}

  To avoid the superfluous information, the user needs to set
{\tt DEBUG = false}  and {\tt bShortINF = true}.
  In this case the listing consists of two parts for two
methods of treatment by using the "set A" and "set B"
weighting functions~\cite{SIMUB} 
(in section "Test Run Input and Output" you can see 
the listing with the case of "set B" only).
  The "set B" weighting functions are the linear combinations
of six angular functions which define the amplitude of the process
while the "set A" weighting functions are not expressed linearly
via the angular functions.
  This is the main reason why 
the statistical errors for observables in case of "set B"
is about 2 times smaller than for the "set A".
  In previous sections we have described  the results for "set B" only.

  In previous sections we have used three keys for tuning the program
{\tt BtoVVana}: $\Gamma'$, $T$ and $T_0$.
  To improve solving the equations for $\Delta\Gamma_{L,H}$,
the user can change limits of arguments in the method 
{\tt mu\_DG\_Equa\_Solut} (class {\tt T\_StandardModel\_DG})
which solves the equations.

\section{Conclusion}
  Extraction of physics information
in decays  $B_s^0\to J/\psi\phi$ and $B_d^0\to J/\psi K^*$
by using the angular moments method with time dependent 
and time-integrated observables,
has a number of attractive features, which are demonstrated
by package {\tt BtoVVana}:
\begin{itemize}
  \item it is an unbinned method;
  \item it uses full time informative contents of time-dependent decays;
  \item it uses full available statistics;
  \item it gives stable results in case of small statistics;
  \item it allows one to separate the extraction of physics values;
  \item it allows one to use different scenarios in case of the signals
        beyond Standard Model, or in case when it is justified;
  \item it is a flexible tool because it has a different ways
     to tune the extraction of observables (for example,
     to tune solving the equations) in the process
     of the real data treatment;
  \item and it is a visual method (see, for example,
     the Fig.4 in ~\cite{SIMUB}
     for dependences of $\hat{b_{1,2}}$  on $\Delta\Gamma$).
\end{itemize}

  The detailed tests have been performed for the package 
{\tt BtoVVana} by means of a precise generator {\tt SIMUB}.
  The tests have checked mutually the both programs
  {\tt SIMUB} and {\tt BtoVVana} with high precision.

  The complex {\tt BtoVVana - SIMUB} can be used
to test other methods of extraction of physics information
from decays  $B_s^0\to J/\psi\phi$ and $B_d^0\to J/\psi K^*$.

  The program  {\tt BtoVVana} has clear structure
and can be used as template to include new methods of treatments.

\section*{Acknowledgements}
  This work is dedicated to memory of \fbox{A.A.Belkov} 
  who was the initiator of this work.
\newpage


\newpage

\section*{Test Run Input and Output}

  User defined parameters are defined in section 4 and 
placed in main program:
{\small
\begin{verbatim} 
TROOT root("Program BtoVVana"," xxx ");
int main(int argc, char **argv) {
 TApplication *theApp= new TApplication("App", &argc, argv);
 cout<<" ***** main() Start ***** "<<endl;
//============= user set, Files with events and parameters =====
 const Int_t NFiles = 10; const Char_t *cFile[NFiles];
 cFile[ 0] = "$DAT1/10000ev_B0sJPsiPhi_1.root";
 cFile[ 1] = "$DAT1/10000ev_B0sJPsiPhi_2.root";
 ...
 cFile[ 9] = "$DAT1/10000ev_B0sJPsiPhi_10.root";
 Int_t KF_Code        =531;  // 531 -> B0s,Bbar0s, 511 -> B0d,Bbar0d
 Bool_t DEBUG         =false;// switch for extended listing
 Bool_t bShortINF     =true;
 Int_t  Mode_Loop     =2;    // 2: DGamma measurement; 3: angle distributions
 Double_t Time_UP_mmDc=2.,Time_maximal_mmDc=2.,Time_Zero_mmDc=0.5;//mm/c
 Double_t GammaS_mmDcm1; 
 if    (KF_Code ==531) GammaS_mmDcm1 = 2.278;
 else if(KF_Code==511) GammaS_mmDcm1 = 2.27844;// Gd = 2.12326;
 else out_exit("main: check KF_Code or set GammaS_mmDcm1. STOP.");
 Bool_t   bRealData   = false;
 Double_t ErrFactor=0.2,Reso_CosThetap=0.022,Reso_CosThetaKp=0.0076,
	  Reso_Chi=0.04,Reso_t=0.03; // [Reso_t] = mm/c
 Int_t  NEVENTS = 100000;
 Bool_t bSIMPLE_HISTOGR = false;
//============= end user setting. ================
 T_Physics *Phys = new T_Physics();
 Phys->Init(Time_UP_mmDc, Time_maximal_mmDc, 0., ErrFactor*0.,
    3.14, ErrFactor*3.14, 0., ErrFactor*0., 0.33, ErrFactor*0.33,
    0.33, ErrFactor*0.33, 0.33, ErrFactor*0.33,
    true, GammaS_mmDcm1*Phys->mmDcm1_To_GeV,
    GammaS_mmDcm1*Phys->mmDcm1_To_GeV*1.2, // GL
    ErrFactor*GammaS_mmDcm1*Phys->mmDcm1_To_GeV*1.2, // GL err
    GammaS_mmDcm1*Phys->mmDcm1_To_GeV*0.8, // GH
    ErrFactor*GammaS_mmDcm1*Phys->mmDcm1_To_GeV*0.8, // GH err
    Reso_CosThetap, Reso_CosThetaKp, Reso_Chi, Reso_t);
 if(Mode_Loop==2){
  T_Accum_Measure *AM = new T_Accum_Measure(DEBUG, bShortINF,
     KF_Code, NFiles, cFile, Mode_Loop, 
     Time_UP_mmDc,Time_maximal_mmDc, Time_Zero_mmDc,
     GammaS_mmDcm1, bRealData, Phys);
  AM->Loop(NEVENTS, bSIMPLE_HISTOGR);
 }
 else if(Mode_Loop==3){
  T_run_Read *SRAM=new T_run_Read(KF_Code,NFiles,cFile,Mode_Loop,DEBUG);
  SRAM->Loop(NEVENTS, bSIMPLE_HISTOGR, 0);
 }
 if(Mode_Loop==3) theApp->Run();
 cout<<" ***** main() End ***** "<<endl;
}
\end{verbatim} 
}

  The output listing in short format has the form:
{\small
\begin{verbatim} 
 ***** main() Start ***** 
--- ctor T_Accum_Measure ---
weight func [set B]
--- ctor T_TimeIntObs: ---
  Gamma^prime = 4.49511e-13 GeV = 2.278 [mm/c]^-1
  T_imeUP_mmDc = 1.01355e+13 GeV^-1 = 2 mm/c
  T_imeMAX_mmDc= 1.01355e+13 GeV^-1 = 2 mm/c
[initial condition for other three kind of observables is placed here]
Inform from T_TimeIntObs::Parameter_SIMUB in case of bSIMUB_Test: 
 Gamma_B0s  = 4.49598e-13 GeV = 2.27844  [mm/c]^-1
 DG         =-8.99197e-14 GeV =-0.455689 [mm/c]^-1
 Gp(GL)     = 4.94558e-13 GeV = 2.50629  [mm/c]^-1
 Gm(GH)     = 4.04638e-13 GeV = 2.0506   [mm/c]^-1
 0.5*(Gm+Gp)= 4.49598e-13 GeV = 2.27844  [mm/c]^-1
 (Gm-Gp)= -8.99197e-14 GeV = -0.455689   [mm/c]^-1
 DG/G   = -0.2
 RESOLUTION: dcTb1, dcTa1, dChi, dt = 4e-05, 4e-05, 0.000125664, 4e-05
-------------------------------------------
BEGIN:  T_Accum_Measure::Average_Measure_OutResult 
*** [set B] ***  T_Accum_Measure::Average_Measure_OutResult 
 NEVENT = 100000 TimeUP = 2 mm/c,   GammaS = 2.278 [mm/c]^-1
-------------- BEGIN  TGp  -2prime- BEGIN --- 
o0= 2.1242+/-0.024(  1.1%)+/-0.0000(0%)= 2.1169  = oE0
o1= 1.0584+/-0.033(  3.1%)+/-0.0000(0%)= 1.1760  = oE1
o2= 1.0787+/-0.035(  3.2%)+/-0.0000(0%)= 0.9889  = oE2
o3=-0.0165+/-0.031(188.3%)+/-0.0000(0%)= 0.0099  = oE3
o4=-1.5242+/-0.049(  3.2%)+/-0.0000(0%)=-1.5778  = oE4
o5= 0.0469+/-0.047( 99.8%)+/-0.0000(0%)=-0.0133  = oE5
 From T_Reco_CumObs::Out_Result_CumObs*** 
-------------------------------------------------------------------------
     cVal      Theor    MC  er_MC(erV/V%) er_st_cor(e/eV%) er_phys(e/eV%)
      A_02/A_P2  1.8000  2.0070 0.072(  3.6%) 0.072(  3.6%) 0.00 (  0.0%)
      A_P2/A_02  0.5556  0.4983 0.018(  3.6%) 0.018(  3.6%) 0.00 (  0.0%)
      A_T2/A_02  0.2963  0.3221 0.26 ( 79.5%) 0.012(  3.6%) 0.26 ( 99.9%)
            A02  0.5400  0.5493 0.077( 14.1%) 0.005(  0.9%) 0.077( 99.8%)
            AP2  0.3000  0.2737 0.039( 14.3%) 0.008(  2.9%) 0.039( 97.9%)
            AT2  0.1600  0.1769 0.12 ( 65.5%) 0.006(  3.4%) 0.12 ( 99.9%)
     cos(d2-d1) -1.0000 -1.0166 0.036(  3.6%) 0.036(  3.6%) 0.00 (  0.0%)
cos(d1)*sin(phi) 0.0400 -0.0672 0.14 (201.4%) 0.13 (188.3%) 0.048( 35.5%)
cos(d2)*sin(phi)-0.0400  0.1350 0.17 (122.7%) 0.13 ( 99.8%) 0.096( 58.2%)
-------------- END  TGp  -2prime- END ------------------
[... output for other three observables is placed here]
SM_DG_TGp_TG0  from T_AngMomMethod::Out_Result_TimeIntObs# BEGIN 
-----------------------------------------------------------
 cVal Val_TheorC  Val_MC+/-erVal_MC(erV/V%)+/- er_sys(e/eV%) 
 DGL_0   -0.4566 -0.4493+/-  0.0388(  8.6%)+/-  0.0000 (0%) 
 DGL_1   -0.4566 -0.8170+/-  0.1113( 13.6%)+/-  0.0000 (0%) 
 DGH_2   -0.4548 -0.6876+/-  0.0912( 13.3%)+/- -0.0000 (0%) 
 DGL_4   -0.4566 -0.5358+/-  0.1068( 19.9%)+/-  0.0000 (0%) 
  DG_0   -0.4557 -0.5685+/-  0.0496(  8.7%)+/-  0.0000 (0%) 
  DG_1   -0.4557 -0.7523+/-  0.0720(  9.6%)+/-  0.0000 (0%) 
  DG_4   -0.4557 -0.6117+/-  0.0703( 11.5%)+/-  0.0000 (0%) 
  G_0     2.2784  2.2184+/-  0.0248(  1.1%)+/-  0.0000 (0%) 
  G_1     2.2784  2.3103+/-  0.0360(  1.6%)+/-  0.0000 (0%) 
  G_4     2.2784  2.2401+/-  0.0351(  1.6%)+/-  0.0000 (0%) 
DG_0/G_0 -0.2000 -0.2563+/-  0.0225(  8.8%)+/- -0.0000 (0%) 
DG_1/G_1 -0.2000 -0.3256+/-  0.0316(  9.7%)+/- -0.0000 (0%) 
DG_4/G_4 -0.2000 -0.2731+/-  0.0317( 11.6%)+/- -0.0000 (0%) 
----------------------------------------------------------
[... output for other two types of DG extraction is placed here]
END *** T_Accum_Measure::Average_Measure_OutResult() END ***
####### T_run_Read::Loop End T_run_Read::Loop(). ########
 ***** main() End ***** 
\end{verbatim} 
}

\newpage

\section*{Figure captions}

{\bf Figure 1.} Structure of the package {\tt BtoVVana}.
   Inserted squares mean derived classes.
   Diamond  inside the class means the data members 
   of the classes shown by arrow.

{\bf Figure 2.} Definition of physical angles to describe 
 decays $B^0_s(t),\overline{B}^0_s(t)\to J/\psi(\to l^+l^-)\,\phi(\to K^+K^-)$ 
 in the helicity frame~\cite{SIMUB}.

{\bf Figure 3.}  Angular distributions obtained by means of class 
  {\tt T\_run\_Read} for 100\,000 events with 
  $B_s^0\to J/\psi(\to l^+l^-)\phi(\to K^+K^-)$ decays
 ("main" setting of physics parameters, see section 5).

\newpage
\begin{center}
\epsfysize=0.55\textheight
\epsfbox{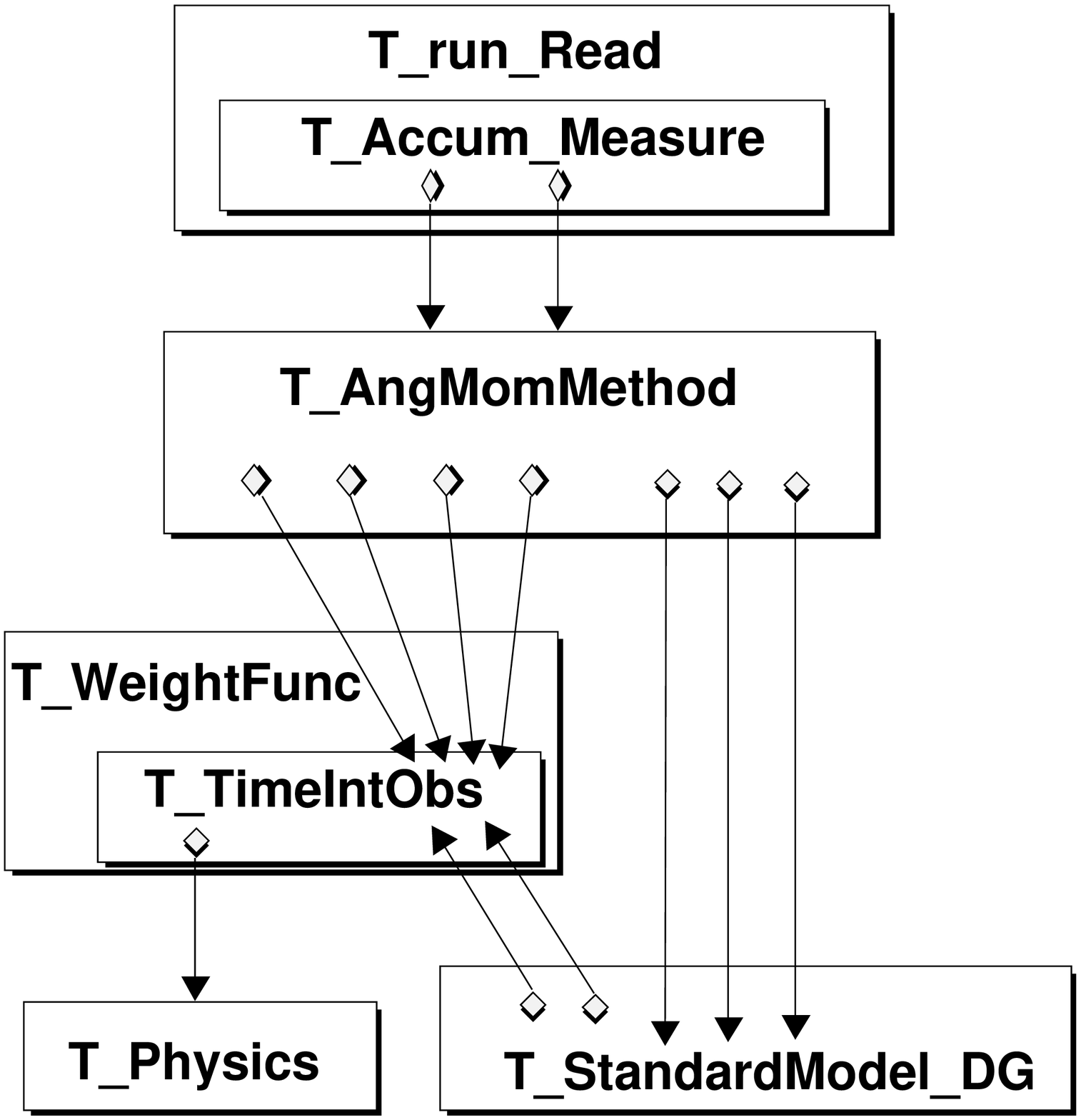}
\end{center}
\vspace{10mm}
{\bf Figure 1}
\begin{center}
\epsfysize=0.55\textheight
\epsfbox{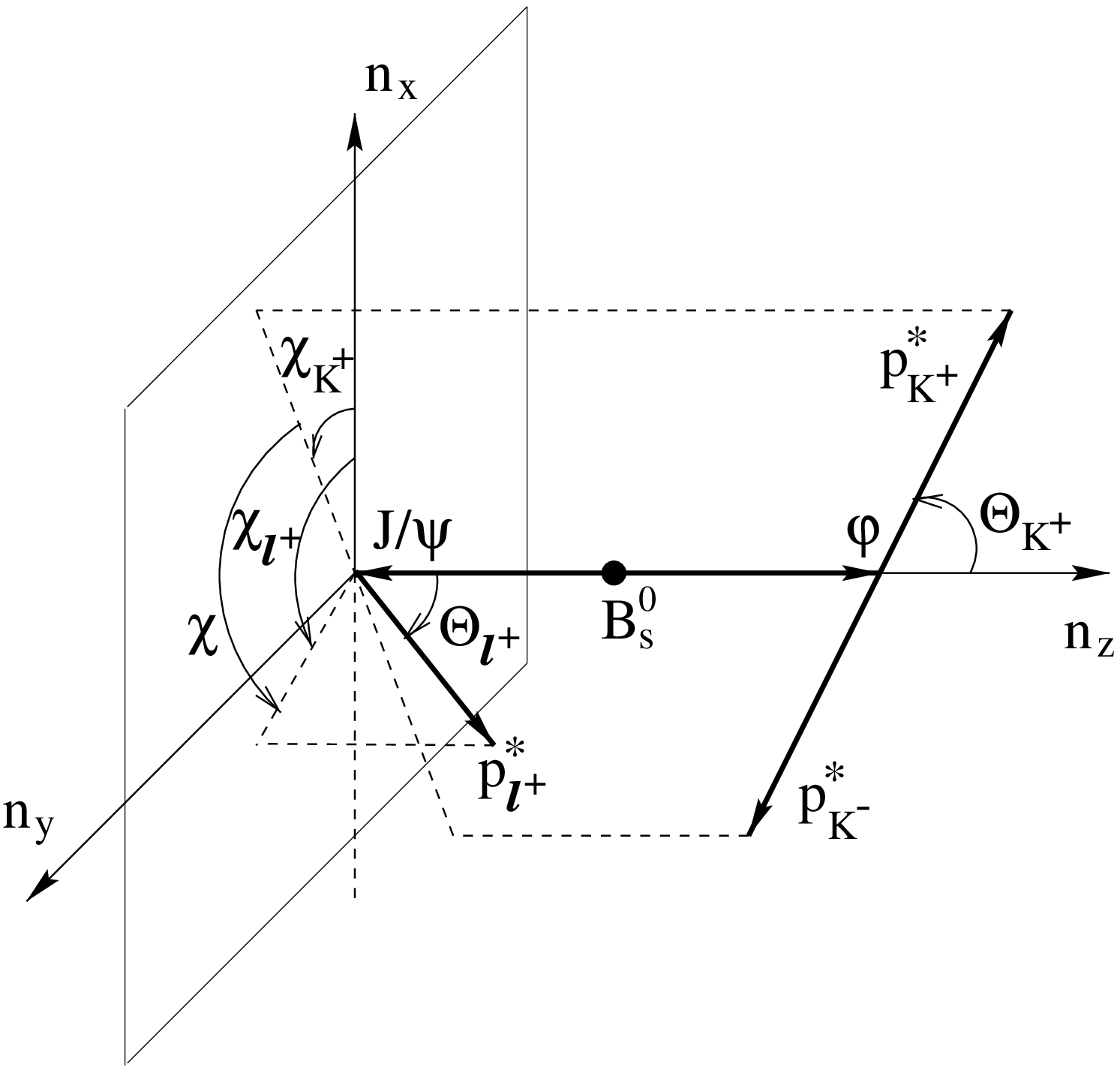}
\end{center}
\vspace{10mm}
{\bf Figure 2}
\begin{center}
\epsfysize=0.55\textheight
\epsfbox{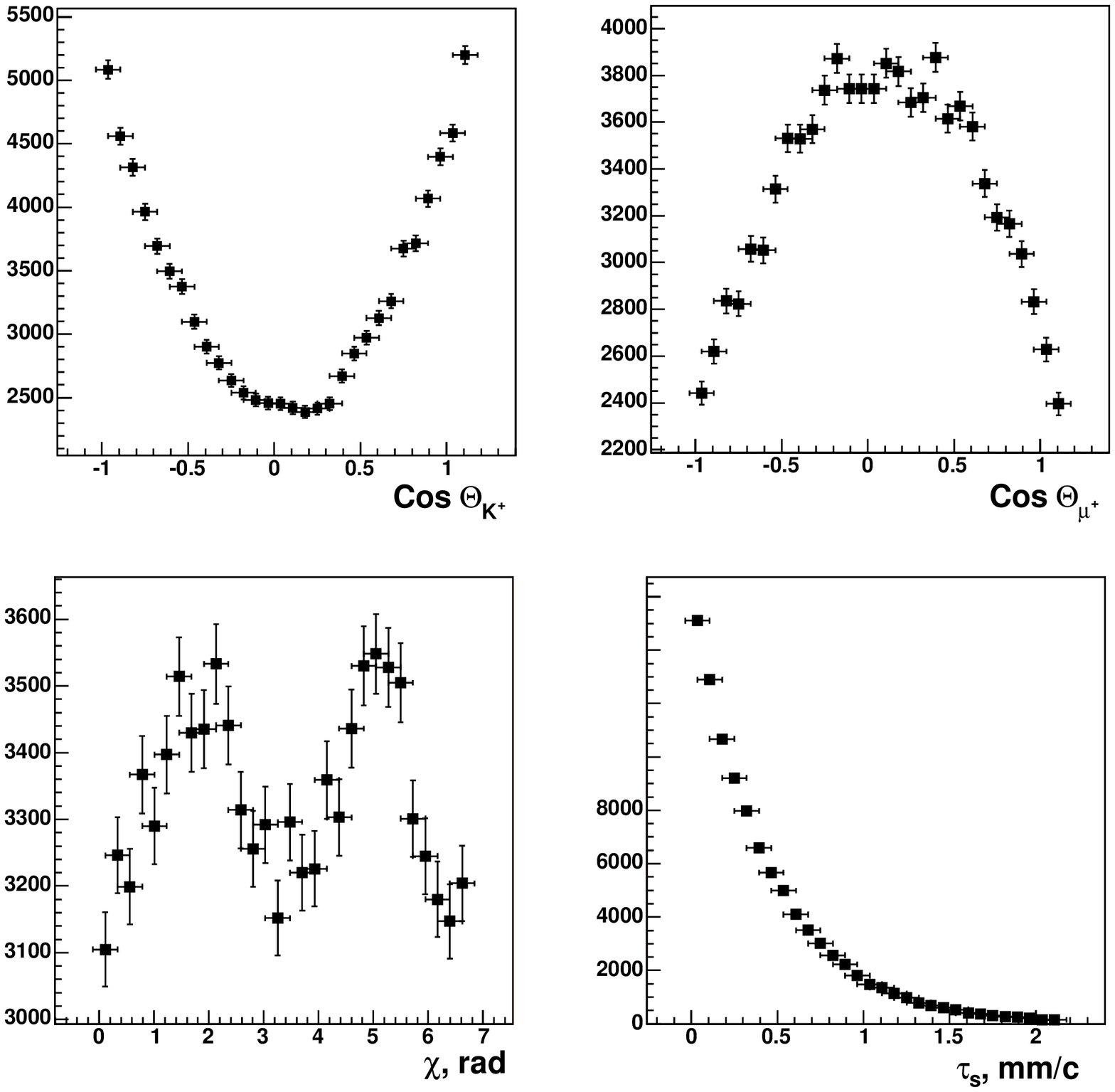}
\end{center}
\vspace{10mm}
{\bf Figure 3}


\begin{thebibliography}{99}
\bibitem{SIMUB} A.~Belkov and S.~Shulga, Comp.Phys.Com. 156 (2004) 221-240;
       A.~Belkov and S.~Shulga, Part. Nucl. Lett., 2[117] (2003) 12, hep-ph/0301105;
       see also  http://cmsdoc.cern.ch/\~\,shulga/SIMUB/SIMUB.html.
\bibitem{337evts} V.M.Abazov e.a. FERMILAB-Pub-04/225-E, hep-ex/0409043 v3 (10 Dec 2004);
       K.Yip, in Proceedings of 39th Rencontres de Moriond on QCD and High-Energy 
       Hadronic Interactions, La Thuile, Italy, 28 Mar - 4 Apr 2004, hep-ex/0405024.
\bibitem{CERN-CKM} Proceedings of the Workshop "The CKM matrix and the unitary triangle",
  13-16 February, 2002, Chapt.3. Editors: M.Battaglia, A.J.Buras, P.Gambino and A.Stocchi,
  CERN-2003-002-corr, 10 October 2003, hep-ph/0304132.
\bibitem{8548evts} K. Abe e.a. BELLE-CONF-0438/ICHEP04 8-0864.
\bibitem{LHC_SMPhys} Proceedings of the Workshop on Standard Model Physics
  (and more) at the LHC. Sect."B decays".
  Editors: G.Altarelli, M.L.Mangano, CERN 2000-004, 9 May 2000.
\bibitem{dighe2} A.S.~Dighe, I.~Dunietz and R.~Fleischer, Eur. Phys. J. 
                 C6 (1999) 647.
\bibitem{ROOT} R.~Brun and F.~Rademakers,  Nucl. Instrum. Meth. A389 (1997) 
               81; see also  http://root.cern.ch/.
\end{thebibliography}
\end{document}